\documentclass[journal=jacsat,manuscript=article]{achemso}

\usepackage[version=3]{mhchem} 

\usepackage{graphicx}%
\usepackage{multirow}%
\usepackage{amsmath,amssymb,amsfonts}%
\usepackage{amsthm}%
\usepackage{mathrsfs}%
\usepackage[title]{appendix}%
\usepackage{xcolor}%
\usepackage{textcomp}%
\usepackage{manyfoot}%
\usepackage{booktabs}%
\usepackage{algorithm}%
\usepackage{algorithmicx}%
\usepackage{algpseudocode}%
\usepackage{listings}%
\usepackage{ragged2e} 
\usepackage[labelfont=bf, justification=justified, singlelinecheck=false]{caption}
\usepackage{caption}
\usepackage{subcaption}
\usepackage{float}
\usepackage{appendix}
\usepackage{dcolumn}
\usepackage{bm}
\usepackage{ragged2e} 
\usepackage{mciteplus}

\mciteErrorOnUnknownfalse

\author{Jiawei Zhang}
\affiliation[Southeast University]{School of Physics, Southeast University, Nanjing, 211189, China}

\author{Andong Liu}
\affiliation[The Chinese University of Hong Kong]{Department of Physics, The Chinese University of Hong Kong, Shatin, N.T., Hong Kong}

\author{Jin Wang}
\affiliation[Southeast University]{School of Physics, Southeast University, Nanjing, 211189, China}

\author{Zhenggao Dong}
\email{zgdong@seu.edu.cn}
\affiliation[Southeast University]{School of Physics, Southeast University, Nanjing, 211189, China}

\title[An \textsf{achemso} demo]
  {Topology of Far-field Signals for Photonic Crystal Slabs}

\abbreviations{IR,NMR,UV}

\makeatletter

\newcommand{\Rmnum}[1]{\expandafter\@slowromancap\romannumeral #1@}
\makeatother
\begin{document}


\begin{abstract}
The study of band topology in photonic crystals was primarily focused on near-field effects, including edge states and high-order corner states. However, this work investigated the polarization distribution of radiated fields for photonic crystal slabs to get their far-field properties of band topology. We introduced a new topological invariant—the winding number of far-field polarization around the Brillouin zone boundary and confirmed a robust correspondence between it and the Chern number of energy bands from the perspective of symmetry, which can be used to analyze the process of topological phase transition. It is found that changes in the winding number and Chern number, associated with the exchange of far-field polarization singularities, especially for bound states in the continuum(BIC), will emerge during phase transition. These findings offer new insights for further understanding the intriguing properties of topological materials.
\end{abstract}

\section{Introduction}\label{sec1}
The discovery of the quantum Hall effect introduced the concept of topology into condensed matter physics \cite{ref1}, and it was developed into the study of photonic crystals rapidly after that\cite{ref2}. The one-dimensional Su-Schrieffer-Heeger model was realized in the optical platforms \cite{ref3} and two-dimensional states, including the quantum spin Hall effect \cite{ref4}, the quantum valley Hall effect \cite{ref5}, and high-order topological corner states \cite{ref6}, were also studied through optical platforms. Three-dimensional Weyl points serving as magnetic monopoles of Berry curvature in momentum space and the Fermi arcs connecting Weyl points with different chiralities in optical systems have also attracted widespread interest \cite{ref7,ref8}. Numerous studies on the nodal lines in photonic crystals have been reported \cite{ref9}, and the topological properties of photonic crystals were also investigated through synthetic dimensions \cite{ref10}. Among these researches, both theoretical and experimental investigations have been extensively carried out \cite{ref11}, but they were limited to near-field effects originating from the topology of energy bands. The question of whether the topology of bands has far-field effects naturally emerged.
Some researchers have attempted to explore it \cite{ref12,ref13}, while clear theoretical correspondence between them still needs to be provided. This work rigorously derives the correspondence between the Chern number and the integral of the far-field polarization angle around the boundary of the Brillouin zone(winding number) in photonic crystal slabs. Surprisingly, the defined winding number could be applied to describe topological phase transition, similar to the Chern number.

The far-field effects in photonic crystal slabs have been extensively studied in recent years \cite{ref14}, among which the bound states in the continuum (BIC) are the most intriguing since they possess extremely high-quality factors \cite{ref15,ref16,ref17}. In addition to BIC, the singularities of far-field polarization, such as C-points, L-lines \cite{ref18,ref19,ref20}, and phase singularities \cite{ref21} have also been widely investigated. The far-field signals of edge states were also studied \cite{ref22}, but they are limited to phenomenological observations or qualitative theories. 
So far, no studies have rigorously revealed the relationship between the far-field singularities mentioned above and the band topology. The winding number presented in this work, representing the sum of the global far-field polarization topological charges, can effectively bridge them, and it can be linked to the Chern number through symmetry, as illustrated in Figure 1.
This study will provide insights for the research of non-Hermitian topology \cite{ref23,ref24}, topological acoustics with far-field signals \cite{ref25}, as well as the study of optical metamaterials \cite{ref26,ref27}. 

 In this work, the correspondence between the invariant (Chern number) of topological bands and the integral of the far-field polarization angle around the Brillouin zone (BZ) boundary (winding number) is proposed firstly based on symmetry analysis. The correspondence is applied to describe the topological phase transition in photonic crystals \cite{ref28,ref29}. 
The topological charge of BIC at the $\Gamma$ point for the hexagonal lattice model is calculated with temporal coupled mode theory \cite{ref30}, and it's found that the Chern and winding numbers will change during the topological phase transition, associated with the exchange of BIC simultaneously. Moreover, In the study of multi-layered photonic crystal slabs, the winding number of the far-field polarization states is found to be conserved when the gap remains open. After that, the far-field properties of the edge states are obtained by solving the Dirac equation. The Chern number is found to be constrained by symmetry, similar to the fact that the high-symmetry points of a system will constrain the topological charge of BIC \cite{ref31}, and it will provide insights into topological phase transition. Finally, the far-field properties under non-Abelian connections are also discussed. The numerical calculation method in Ref. \cite{ref32} is applied to compute the Chern numbers in the work, and the software COMSOL is used to verify theoretical predictions.

\begin{figure}[htbp]
    \centering
    \includegraphics[width=.65\linewidth]{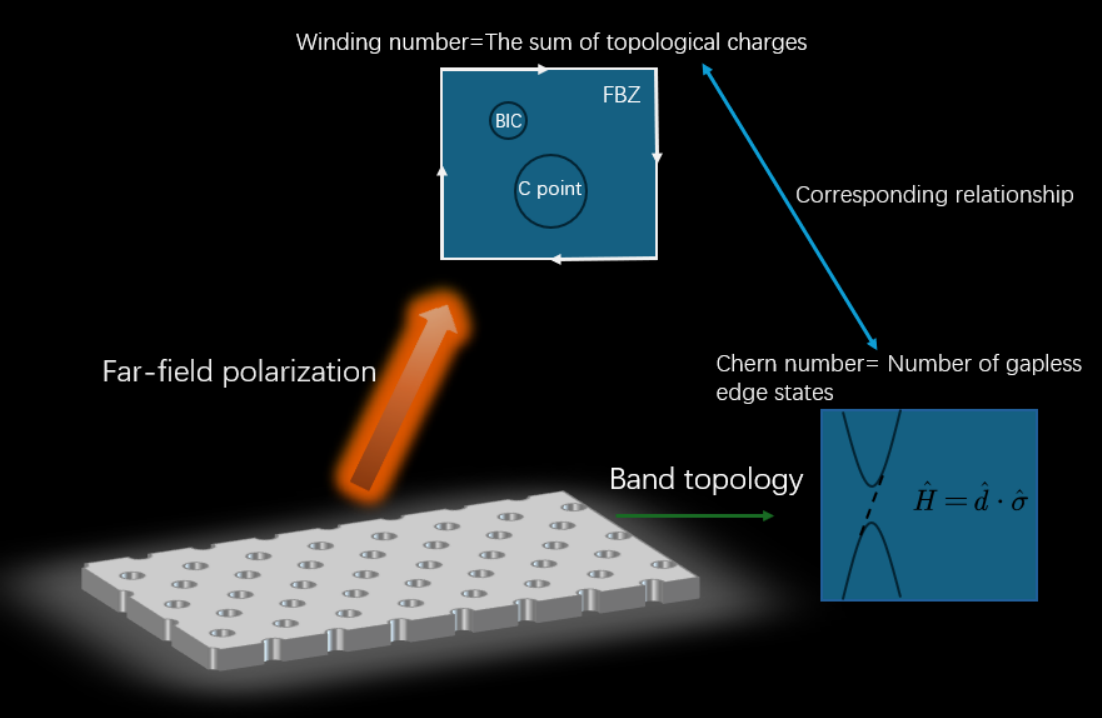}
    \captionsetup{justification=justified, singlelinecheck=false}
    \caption{\label{fig:epsart} Schematic depicting the relationship between the Chern number and the winding number of far-field polarization states in band topology.}
\end{figure}

\section{Results and discussion}
To explore the far-field signals of topological bands from the perspective of symmetry, it is natural to consider that the eigenstates of the Hamiltonian $\phi _{n}\left(r\right)$ can be transformed with the following form:
\begin{equation}
\phi_{n}\left(\Lambda^{-1} r\right)=\sum_{m} \hat{R}(\Lambda)_{n m} \phi_{m}(r).
\end{equation}

For photonic crystals, its master operator can be defined as $\Theta_{\mathbf{k}}=\frac{1}{\epsilon}(\nabla+i \mathbf{k}) \times(\nabla+i \mathbf{k}) \times$\cite{ref30}, and its eigenstates satisfy the transformations described in the equation(1). In the far-field diffraction limit, the polarization could be described with the method proposed in Refs. \cite{ref15,ref31}:
\begin{equation}
\begin{aligned}
	\vec{c}\left( k_x,k_y \right) &=\left( c_x,c_y,c_z \right)\\
	&=\iint_{\text{cell\,\,}}{e}^{ik_xx+ik_y}\vec{E}\left( k_x,k_y \right) \text{d}x\text{d}y/\iint_{\text{cell\,\,}}{\text{d}}x\text{d}y.\\
\end{aligned}
\end{equation}

Since rotations in reciprocal space can be transformed into rotations in real space, the projection of far-field polarization onto the plane of the photonic crystal slab will follow the transformation of the vector representation. With the mathematical calculations in the discussion of Supplementary Information Section \Rmnum{1}, where the vector field of the polarization states is transformed into a complex scalar field, 
it is found that because the far-field polarization and eigenstates of the master operator carry different linear spaces of the same representation, there exists a correspondence between the Chern number and the winding number of the far-field polarization:
\begin{equation}
C=q+\oint_{B Z} i \Gamma^{*}(k)\left(\nabla_{k} \Gamma(k)\right) d k / 2 \pi,
\end{equation}
where $C$ is the Chern number and $q$ is the winding number of far-field polarization, $\Gamma(k)$ is a similarity transformation matrix. Since band crossings require that the eigenstates of a Hamitonian should carry different representations \cite{murakami2007phase} and the integral part in equation (3) will not change by an integer when the gap remains, the winding number of the far-field polarization is a topological invariant, serving as a topological charge for far-field signals. 
\begin{figure}[htbp]
\centering
    \begin{subfigure}[t]{0.4\linewidth}
    \centering
	\captionsetup{justification=raggedright, singlelinecheck=false, labelformat=empty, skip=0pt, position=top}
	\caption*{(a)}
        \includegraphics[width=.85\linewidth]{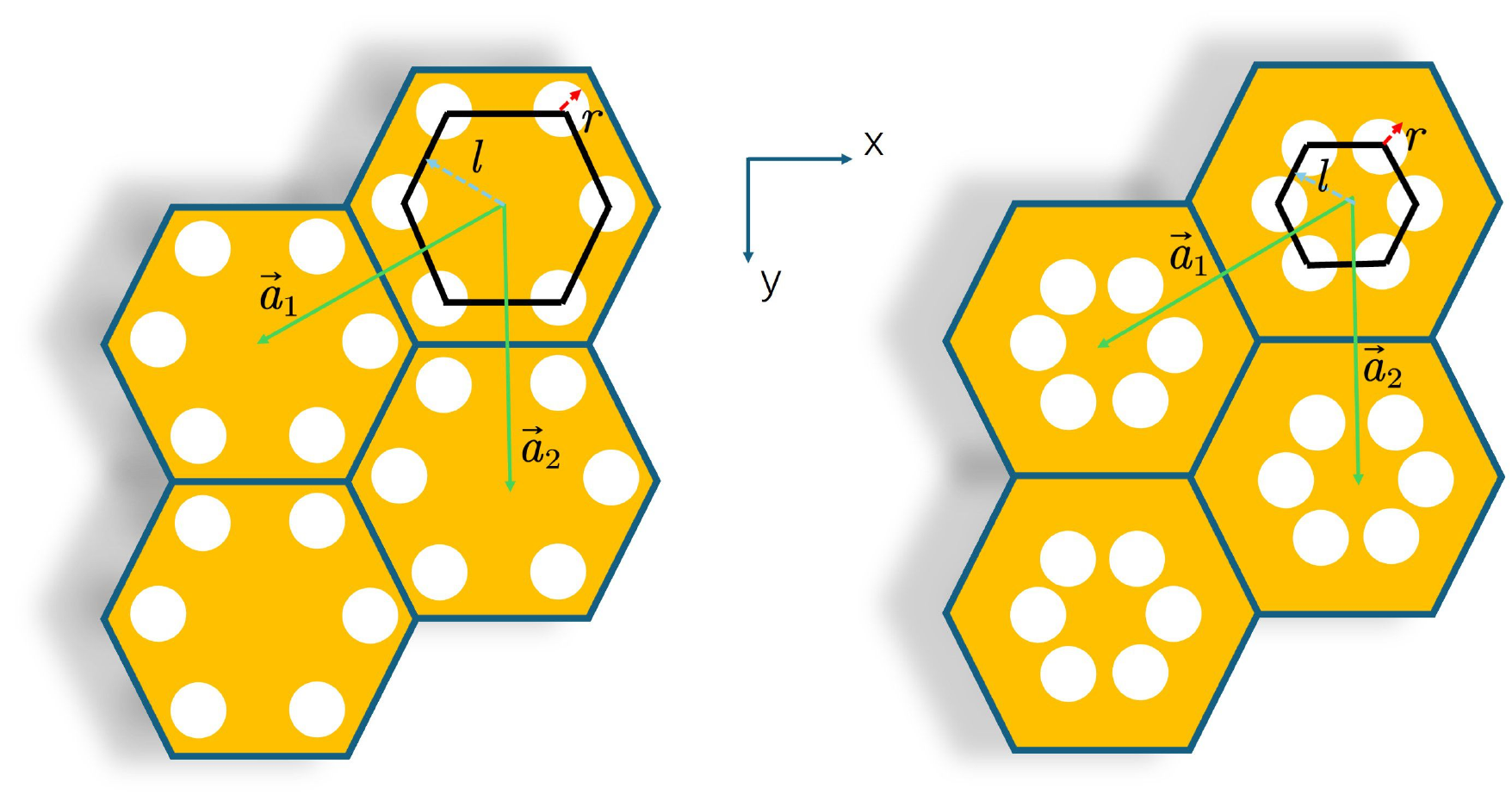}
	\label{fig:sub1}
    \end{subfigure}
    \begin{subfigure}[t]{0.4\textwidth}
    \centering
        \captionsetup{justification=raggedright, singlelinecheck=false, labelformat=empty, skip=0pt, position=top}
        \caption*{(b)}
        \includegraphics[width=.8\linewidth]{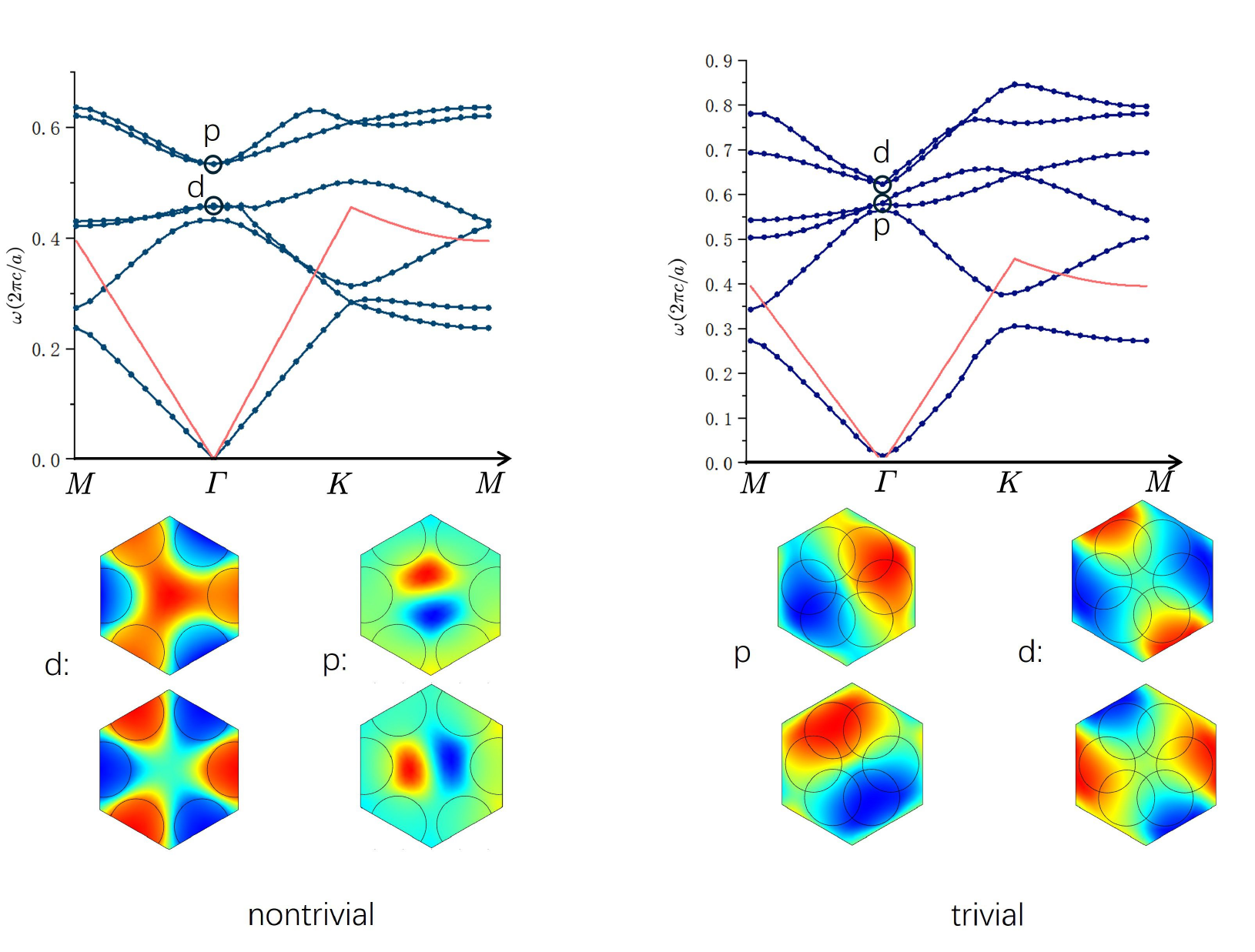}
        \label{fig:sub2}
    \end{subfigure}
   
    \begin{subfigure}[t]{.91\textwidth}
    \centering
        \captionsetup{singlelinecheck=false, labelformat=empty, skip=0pt, position=top}
        \caption*{(c)}
        \includegraphics[width=.9\linewidth]{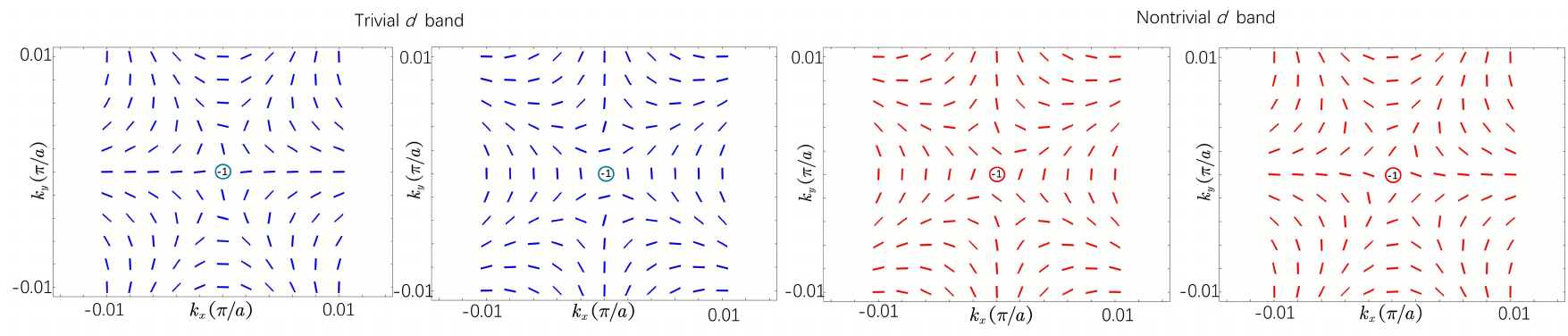}
        \label{fig:sub1}
    \end{subfigure}
\hfill

	        \captionsetup{justification=justified, singlelinecheck=false}

\caption{\label{fig:epsart}\justifying  (a) Schematics of a photonic crystal slab with holes. The image on the left stands for the expanded structure of the hexagonal photonic crystal slab with $a/l < 3$, where $a$ is the lattice constant and $l$ is the length indicated in the figure. When the radius of holes $r=0.2a$, this structure corresponds to a nontrivial topological case. The image on the right stands for a contracted structure of a hexagonal photonic crystal slab, where $a/l > 3$ and $r=0.2a$, which is topologically trivial. (b) Energy bands and the z-component of magnetic field distributions for nontrivial and trivial topological photonic crystal slabs. The subfigures on the top left and top right are the energy bands of trivial and nontrivial topological slabs respectively, where the $p$ mode and the $d$ mode are exchanged after the band closes. The red line represents the light cone. The subfigures on the bottom left and bottom right show the distribution of the z-component of the magnetic field for the $d$ mode and $p$ mode in the nontrivial and trivial slabs, respectively. (c) Left: For a trivial topological slab, two $d$ mode states near the $\Gamma$ point carry a topological charge of -1. The solid blue lines represent the long axis of the far-field polarization. Right: For a nontrivial topological slab, two $d$ mode states near the $\Gamma$ point carry a topological charge of -1, too. The solid red lines represent the long axis of the far-field polarization}
\end{figure}

To describe the process of a topological phase transition in a hexagonal lattice with far-field polarization, a photonic crystal slab depicted in Fig. 2(a) is applied to obtain the band gap of TE modes, which consist of circular holes with a radius of $r$ in a dielectric slab with a permittivity of 10.2. The photonic crystal is immersed in material with a refractive index of 1.46 to generate more modes within the light cone. By varying the radius $r$ of the circular hole and the geometrical length $l$ in a unit cell, different topological phases can be achieved. For example, when $l = 0.275a$ and $r = 0.2a$, it corresponds to a topologically trivial phase, and when $l = 0.475a, r = 0.2a$, it corresponds to a topologically nontrivial phase. We consider a photonic crystal slab with a thickness of 700 nm, and the band structures of TE mode for the topologically trivial and nontrivial modes are shown in Fig. 2(b).

In the case of the trivial phase, it's observed that the lower band at the $\Gamma$ point exhibits a doubly degenerate $p$ mode. Due to the odd parity of the magnetic field, there is a strong coupling between the electric field and the far field, resulting in a lower quality factor. Conversely, the doubly degenerate $d$ mode in the upper band exhibits the opposite behavior, where they completely decoupled in the far field at the $\Gamma$ point due to its symmetry, leading to the divergence of the quality factor and the emergence of a BIC. By applying the temporal coupled mode theory in Supplementary Information Section \Rmnum{2} and analyzing the far-field polarization near the $\Gamma$ point shown in Fig. 2(c), it is observed that for the trivial phase, both the upper bands have a BIC with a topological charge of -1 at the $\Gamma$ point. Similarly, for the nontrivial phase, a topological charge of -1 at the $\Gamma$ point for both lower bands is observed. A Dirac point at the $\Gamma$ point will emerge during the topological phase transition. After the gap is reopened, the bands of $p$ mode and $d$ mode will reverse, and the BIC will be exchanged between bands, as shown in Fig. 3. The Chern number and the winding number of polarization states will change simultaneously. It is worth noting that for crystals with $C_{6v}$ symmetry, considering the tight-binding case and using two $p$ modes and two $d$ modes as a basis, its Hamiltonian can be derived using $kp$ perturbation theory [12]:
\begin{equation}
\hat{H}=\left(\begin{array}{cc}
\hat{H}^{+} & 0 \\
0 & \hat{H}^{-}
\end{array}\right),
\end{equation}
where $\hat{H}^{ \pm}=\hat{d}^{ \pm} \cdot \hat{\sigma}=\left(\begin{array}{cc}
w_{p}+\beta k^{2} & \alpha\left( \pm k_{x}-i k_{y}\right) \\
\alpha\left( \pm k_{x}+i k_{y}\right) & w_{d}-\beta k^{2}
\end{array}\right)$. It is a continuous model rather than a lattice model when calculating the Chern number of the Hamiltonian:
\begin{equation}
C=\iint dk^2\frac{1}{2}\frac{\partial\hat{d}}{\partial k_x}\times\frac{\partial\hat{d}}{\partial k_y},
\end{equation}
where the integration is over an infinite plane rather than the BZ with periodic boundary conditions, which transform the closed manifold into an open plane, resulting in the mapping from $T2$ to $S2$ being reduced to a mapping from the plane to half of $S2$. So the calculated Chern number is not the true Chern number. However, the correspondence between the Chern number and the winding number of the Hamiltonian still exists. For convenience, the winding numbers are acquired by considering the far-field polarization angles of six $M$ points in the BZ boundary.

The corresponding Chern number and winding number before and after a topological phase transition are shown in Fig. 3. It can be observed that the polarization states for the two energy bands of the $p$ mode in the trivial phase have winding numbers of +2 and +1, respectively. After the topological phase transition, they acquire a topological charge of -1, leading to the winding numbers of the lower energy bands becoming +1 and 0. Similarly, for the $d$ modes in the trivial phase, their winding numbers are +1 and -1. After the transition, one topological charge of -1 is lost, resulting in winding numbers of +2 and 0. Therefore, topological phase transition can be characterized by far-field polarization. When the energy gap closes and reopens, the upper and lower energy bands will exchange BIC, changing their Chern and winding numbers. 
\begin{figure}
    \centering
    \includegraphics[width=.85\linewidth]{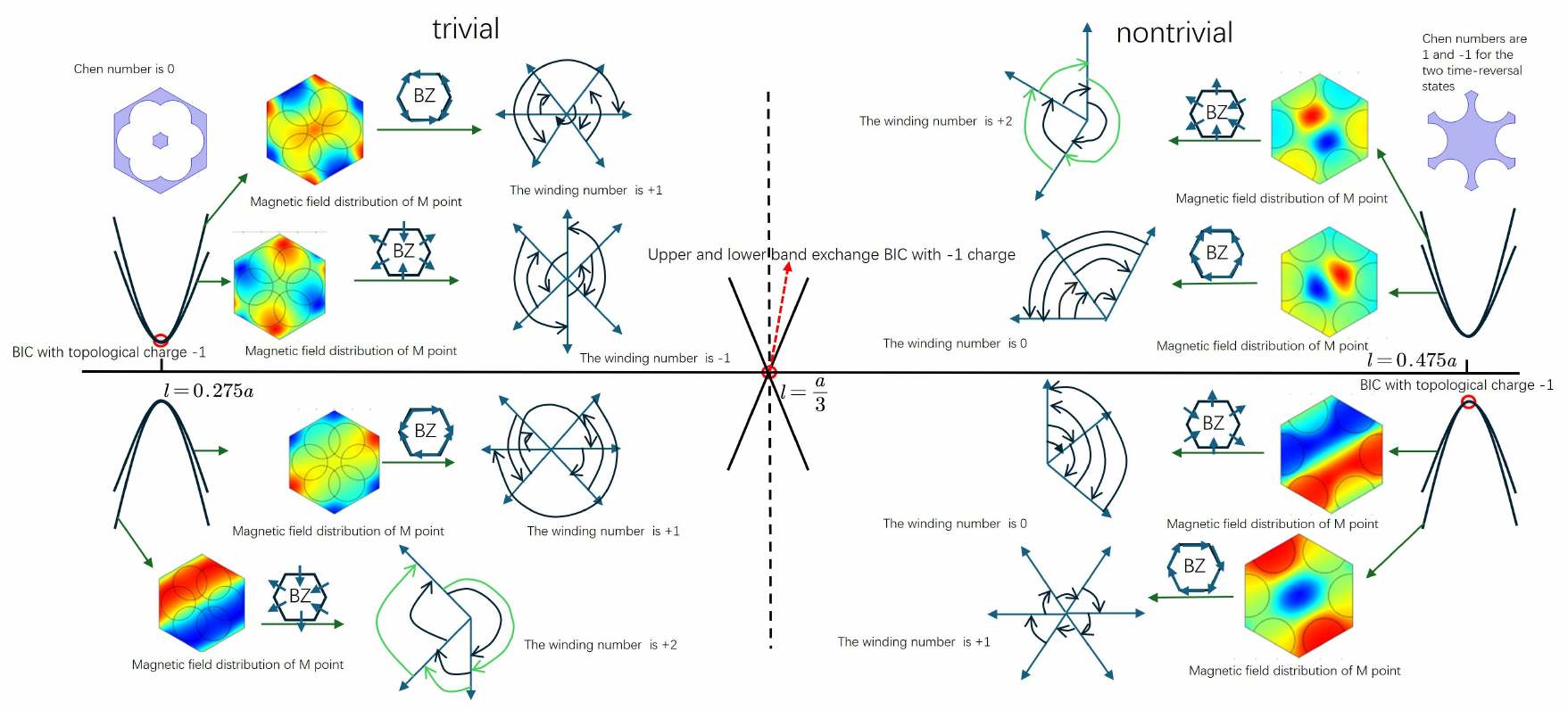}

    \caption{\label{fig:epsart}\justifying  The mechanism of exchanging far-field singularity during topological phase transition.
For the topologically trivial phase, a BIC with a topological charge of -1 (indicated by a red circle) exists at the $\Gamma$ point for the two upper energy bands. The magnetic field for the $M$ points of the energy band is also presented. By analyzing the variation of far-field polarization direction for each $M$ point at the boundary of BZ, it is found that the winding numbers of the far-field polarization for these two energy bands remain 1 and -1 before band inversion. Similarly, the winding numbers for the two lower energy bands are +1 and +2, and there is no BIC at the $\Gamma$ point due to the symmetry of the $p$ mode; after the closure (at $l=a/3$) and reopening of the energy gap, BIC is exchanged between the upper and lower bands, resulting in winding numbers of 2 and 0 for the upper bands and winding numbers of 0 and 1 for the lower bands.}
\end{figure}
For adiabatic transformations, research has shown [18] that the generation or elimination of BIC is accompanied by the disappearance and appearance of C points without changing the winding number. To apply the concept of winding number to multi-layer slabs, we introduced a three-layered slab structure to induce energy gaps and verify that the winding number of far-field polarization is a topological invariant for the energy band after split. When the three crystal slabs shown in Fig. 4(a) are brought close together, each energy band will split into three bands due to the couple between bands. The Hamiltonian of the system can be written as:
\begin{equation}
\hat{H} =\sum_{d=1}^{D}|d><d|\otimes\hat{H}_{ld}+(C\sum_{d=1}^2|d+1><d|\otimes\hat{I}+c.c),
\end{equation}
where we only consider the coupling between the nearest neighboring slabs, and the coupling strength $C=\langle d|\hat{H}| d+1\rangle=C(k)$  increases as the distance decreases. $\hat{H}_{l d}$ is the Hamiltonian for the $d$-th slab, satisfying $H_{l d}|d\rangle=E(k)|d\rangle$. The dispersion relation of $E(k)$ for each crystal slab and the whole system is shown in Fig. 4(b). Calculations in Supplementary Information Section \Rmnum{3} show that while changing the coupling coefficient $C$, the black band in Fig. 4(b) splits into three bands. The winding number of the red band, which emerges after the split of degeneration, could be acquired as below:
\begin{equation}
q'=q_2+\oint dk\nabla_k\frac{1}{2}c_2^2=q_2,
\end{equation}
where $q'$ is the winding number of the red band after the split, $q_2$ is the winding number of the red band before the split, $c_2=\sqrt{\frac{2|C(k)|}{C^{2}(k)}}$. It can be observed that the winding number remains unchanged in the presence of an energy gap. However, when there is symmetry breaking and energy gap closing, this method cannot be used to compute the winding number (Supplementary Information Section  \Rmnum{3}).
\begin{figure}[htbp]
\begin{subfigure}[t]{0.3\linewidth}
\centering
\captionsetup{justification=raggedright, singlelinecheck=false, labelformat=empty, skip=0pt, position=top}
\caption*{(a)}
\includegraphics[width=.9\linewidth]{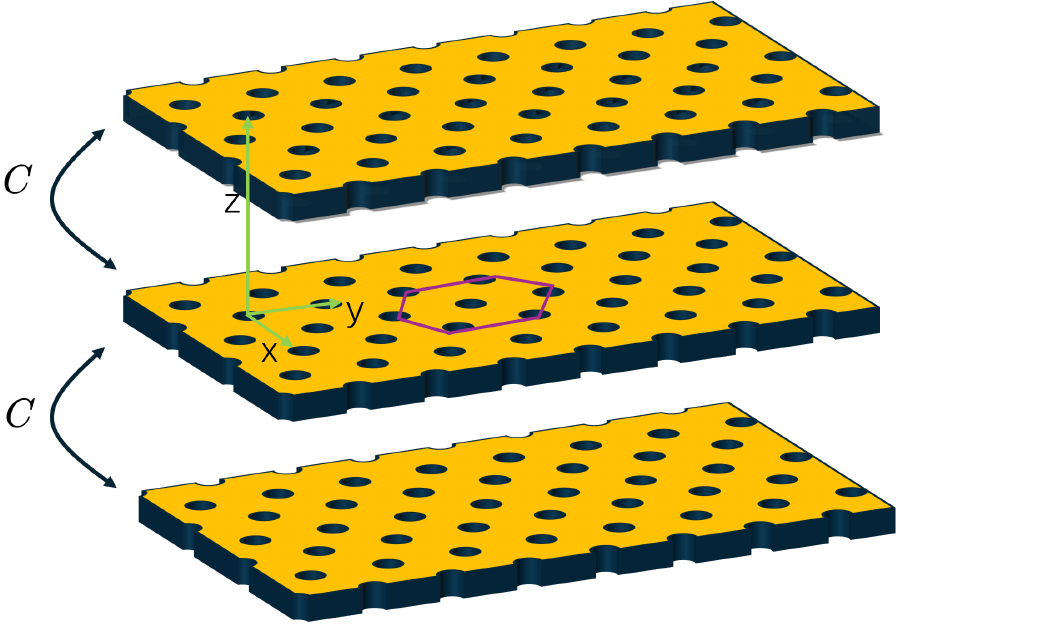}
  \label{fig:sub1}
 \end{subfigure}
 \begin{subfigure}[t]{0.6\textwidth}
 \centering
  \captionsetup{justification=raggedright, singlelinecheck=false, labelformat=empty, skip=0pt, position=top}
  \caption*{(b)}
\includegraphics[width=.95\linewidth]{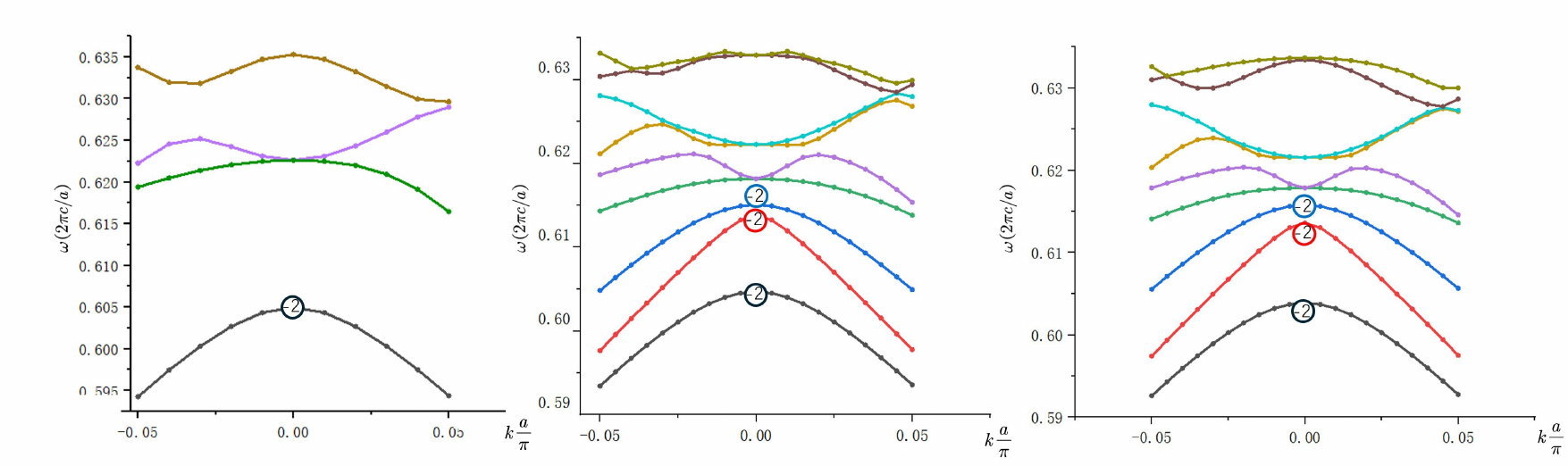}
  \label{fig:sub1}
 \end{subfigure}
 \hfill

 \begin{subfigure}[t]{0.6\textwidth}
  \captionsetup{justification=raggedright, singlelinecheck=false, labelformat=empty, skip=0pt, position=top}

  \caption*{(c)}
\includegraphics[width=.9\linewidth]{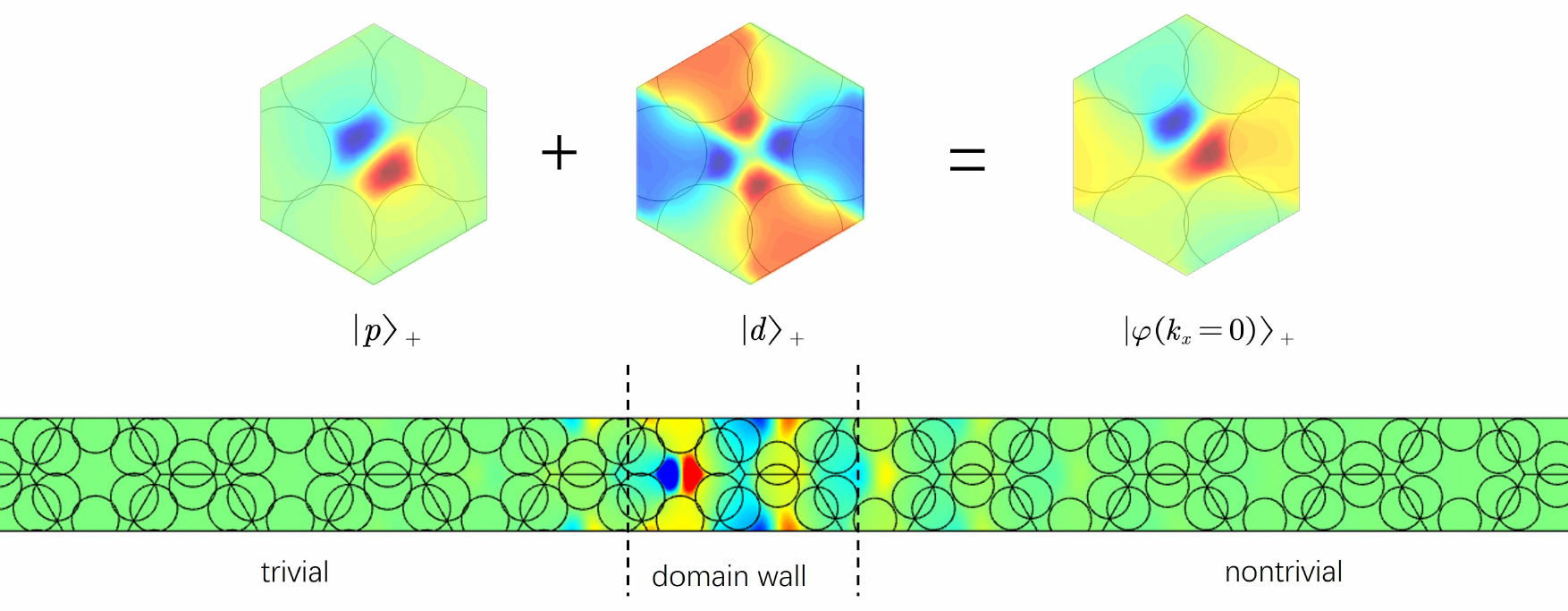}
  \label{fig:sub1}
 \end{subfigure}
 \begin{subfigure}[t]{0.3\textwidth}
  \captionsetup{justification=raggedright, singlelinecheck=false, labelformat=empty, skip=0pt, position=top}
  \caption*{(d)}
\includegraphics[width=.95\linewidth]{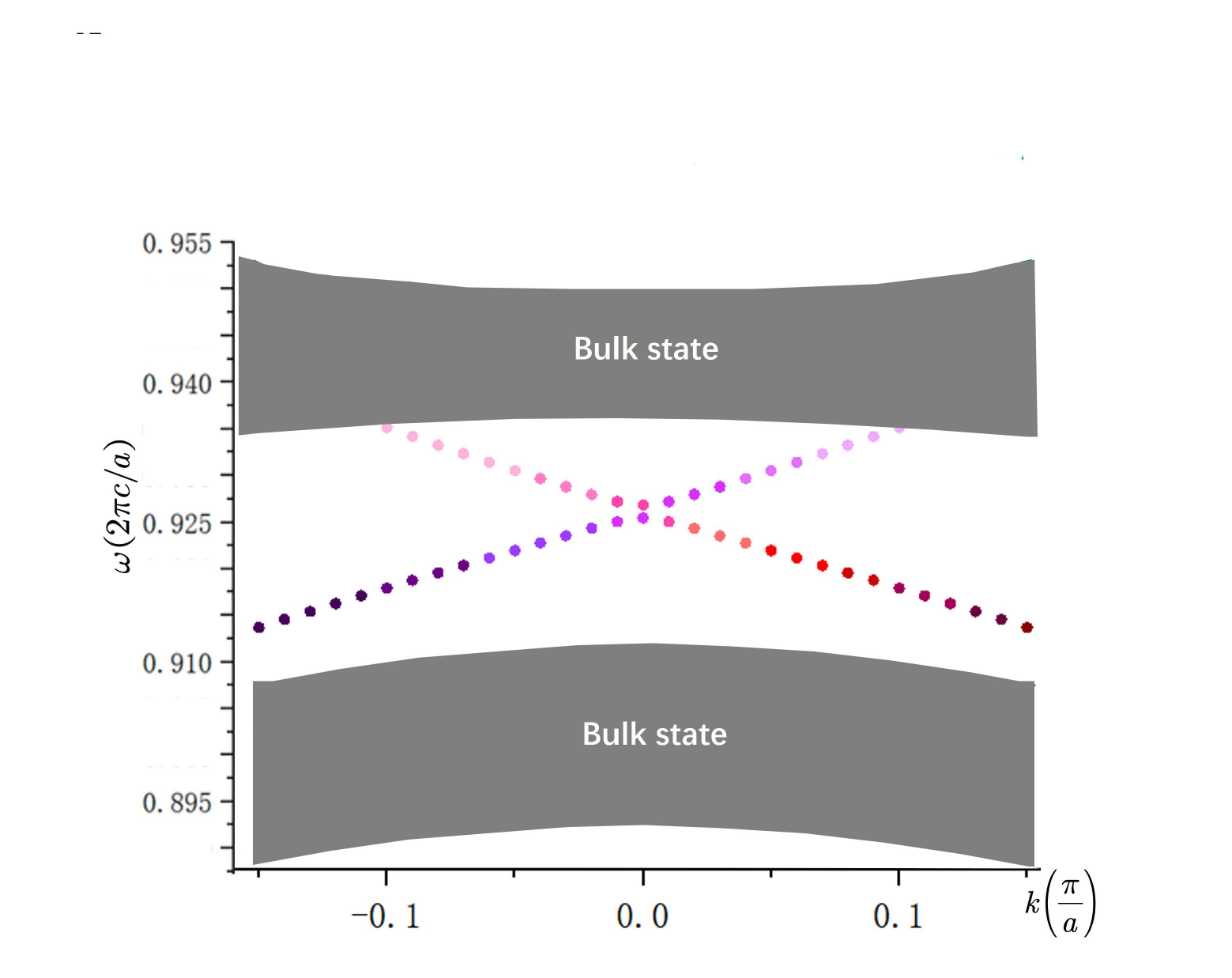}
  \label{fig:sub2}
 \end{subfigure}
\caption{\label{fig:epsart}\justifying (a) Structure of a three-layered crystal slab. When three identical slabs move together from infinity, the energy bands with three-fold degeneracy split due to the overlap of wave functions. Adjusting the distance between slabs can alter the gap of bands. (b) Evolution of the TE mode energy bands of the three-layered slab when changing the distance between slabs. The left diagram corresponds to the band when layers are far apart. The black band carries a -2 BIC charge at the $\Gamma$ point. The middle diagram represents the energy bands when the distance is 108 nm. Due to the interaction between layers, the bands are split, and three bands with a -2 BIC charge also appear at the $\Gamma$ point. The right diagram shows that when the distance becomes 88 nm, the BIC charge at the $\Gamma$ remains unchanged when the energy bands are shifted. (c) The upper part shows the z-component of the magnetic field of edge states for the topologically trivial region, which is obtained by solving the Dirac equation. Since the magnitude of the magnetic field for $p$ mode is relatively large, the distribution of the edge state is dominated by the $p$ mode, which has a lower quality factor. In the topologically trivial region, the thickness of the photonic crystal slab is 100 nm, with other parameters being $l=0.4a$, $r=0.17a$, and $a=1 \mu m$. In the nontrivial region, $r$ is changed to $0.21a$. The distribution of the edge states in the lower part is consistent with the theoretical prediction. (d) Energy bands of a photonic crystal slab with open boundary conditions. The quality factors of the edge state with positive and negative group velocity are indicated respectively with purple and red dots of varying brightness. The darkest shade represents the maximum value of 25.3, and the lightest shade represents the minimum value of 22.2. It can be observed that the modes closer to the lower band have larger quality factors.}
\end{figure}

The above discussion focuses on the far-field property of bulk states; the far-field properties for edge states, which appeared with a domain wall sandwiched between topologically inequivalent materials, should also be studied. Due to the periodicity in one direction, we can obtain the eigen-field of the edge states by solving the following Hamiltonian, which is shown in Supplementary Material Section \Rmnum{4}:
\begin{equation}
\hat{H}^+=\begin{pmatrix}
    u(y) & \alpha(k_x-ik_y)\\
    \alpha(k_x+ik_y) & -u(y)
\end{pmatrix}
\hspace{0.5cm} k_y \xrightarrow{}-i\frac{\partial}{\partial y}.
\end{equation}

Applying the Pauli matrices and considering the linear dispersion relation near the $\Gamma$ point, the Schrödinger equation can be written in the form of the Dirac equation:
\begin{equation}
(u\hat{\sigma}_z+\alpha k_x\hat{\sigma}_x-i\alpha\frac{\partial}{\partial y}\hat{\sigma}_y)\varphi=\nu k_x\varphi.
\end{equation}
The solutions of its edge states are obtained:
\begin{equation}
\varphi=\begin{pmatrix}
    1\\
    \frac{\nu k_x-\sqrt{(\alpha^2-\nu^2)k_x^2+u^2}}{u+\nu k_x}
\end{pmatrix}e^{-\sqrt{(\alpha^2-\nu^2)k_x^2+u^2}|y|/\alpha}.
\end{equation}

From equation (10), it can be observed that for the topologically trivial region ($u>0$), the field is a superposition of the p+ and d+ modes near the $k_x=0$ region, as shown in Fig. 4(c). By analyzing the coefficients in equation (8), it's found that the magnitude of $u$ is determined by the energy gap shown in Fig. 4(d), and the magnitude of $\alpha$ determines the width of bands in the tight-binding model. Likewise, the value of $vk_x$ is between 0 and the magnitude of the gap, indicating that it could create far-field effects. While changing the value of $k_x$, different parity may appear: When $k_x$ equals to $-\frac{u}{\nu}$, the edge state has the characteristics that $|d\rangle_{+} \propto d_{x^{2}-y^{2}}+i d_{x y}$ , carrying a high-quality factor; When $k_x$ satisfies $\nu k_{x} - \sqrt{\left(\alpha^{2}-\nu^{2}\right) k_{x}^{2}+\varepsilon^{2}}=0$ , the edge state satisfies $|p\rangle_{+} \propto p_{x}+i p_{y}$ , carrying a lower quality factor.
In this study, the chosen parameters result in a much larger magnetic field for the $p$ mode than the $d$ mode. Fig. 4(c) shows that their superposition still approximates the $p$ mode, leading to a low-quality factor. In Fig. 4(d), it can be observed that the quality factor of the edge state with negative group velocity becomes larger when increasing $k_x$, while the quality factor will decrease for the edge state with positive group velocity. 
\begin{figure}[htbp]
 \centering
\begin{subfigure}[t]{0.49\textwidth}
    \centering
   \captionsetup{justification=raggedright, singlelinecheck=false, labelformat=empty, skip=0pt, position=top}
    \caption*{(a)}
    \includegraphics[width=.95\linewidth]{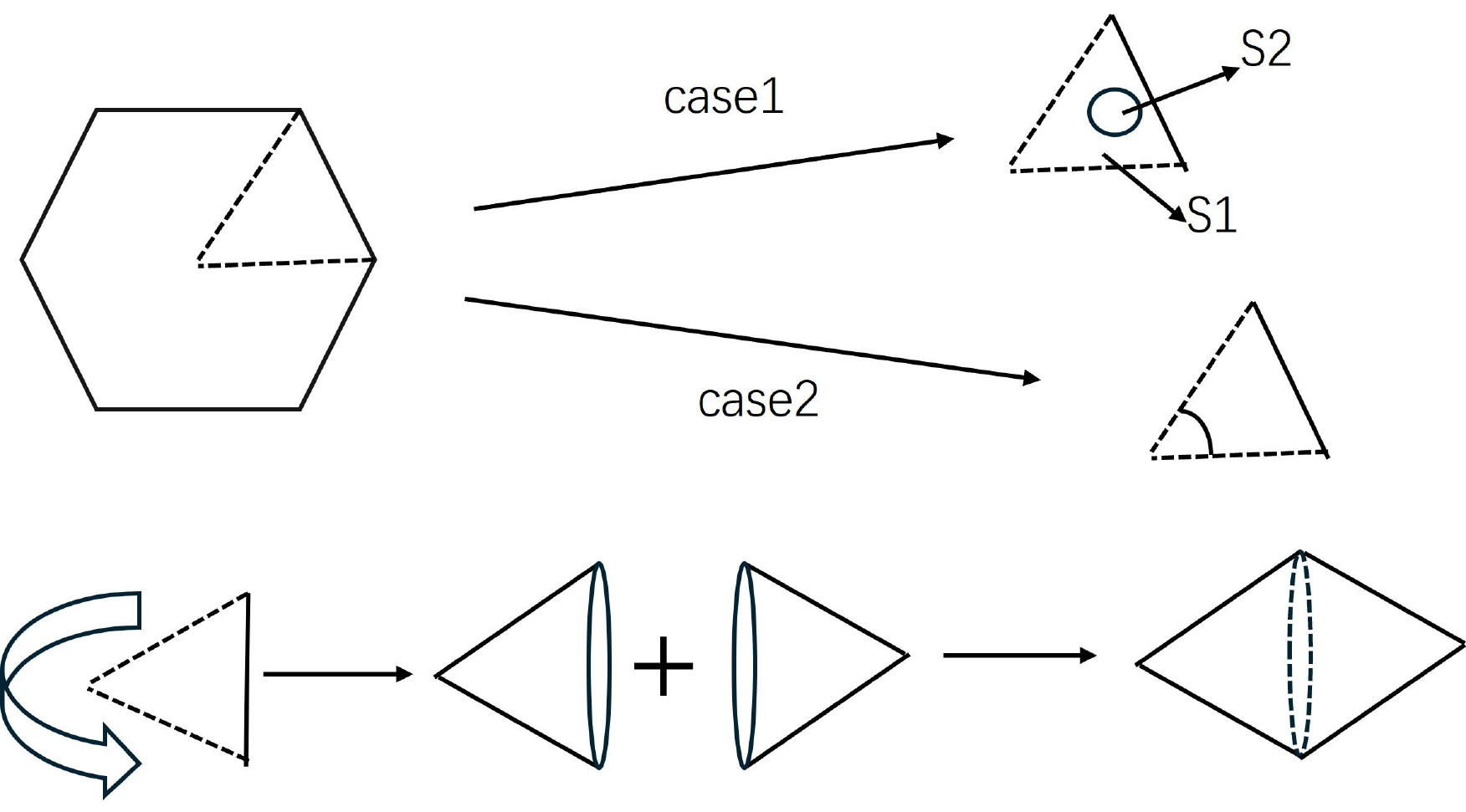}
    \label{fig:sub1}
\end{subfigure}
    \hfill
\begin{subfigure}[t]{0.49\textwidth}
    \centering
    \captionsetup{justification=raggedright, singlelinecheck=false, labelformat=empty, skip=0pt, position=top}
    \caption*{(b)}
    \includegraphics[width=.9\linewidth]{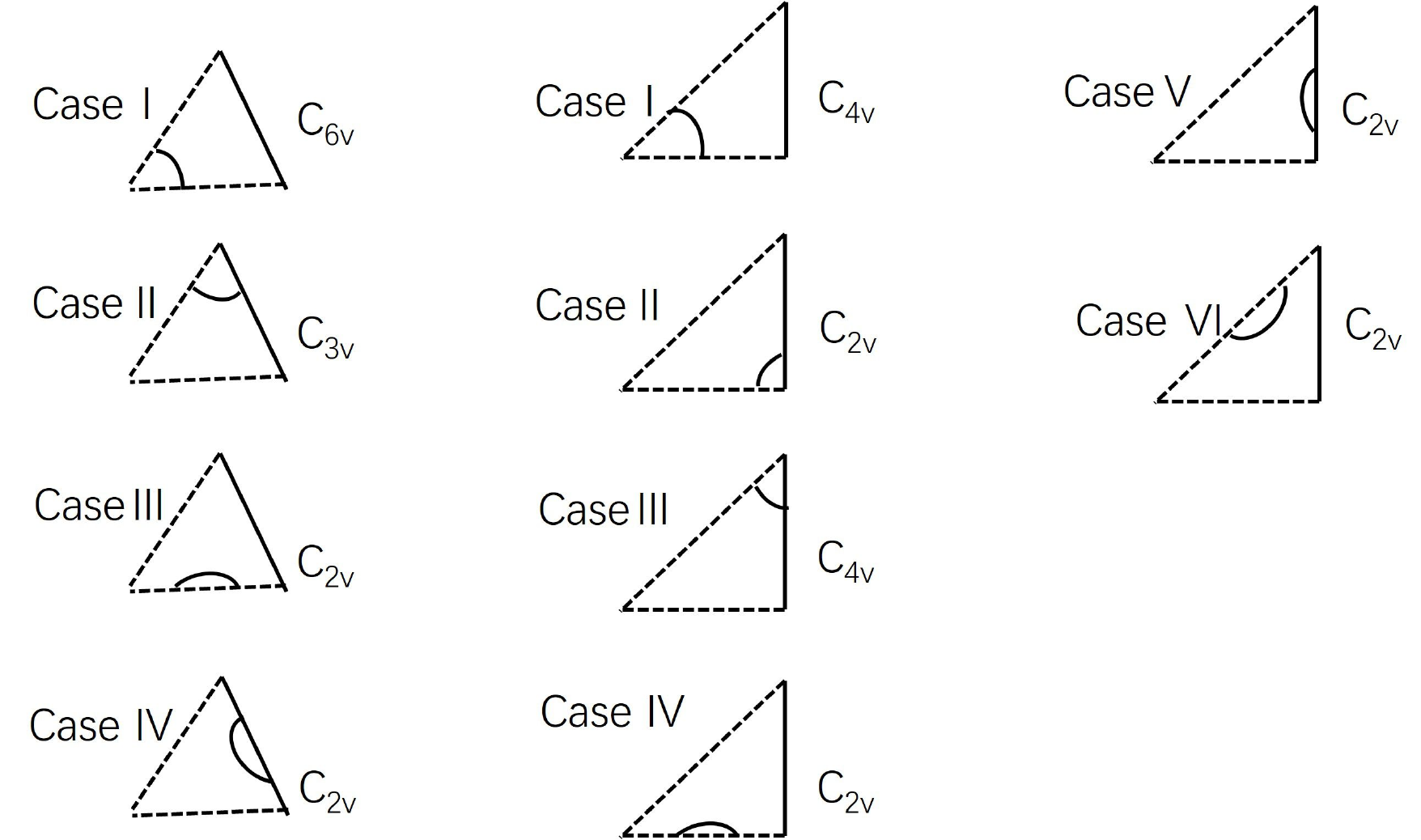}
    \label{fig:sub2}
\end{subfigure}

\caption{\label{fig:epsart}\justifying  (a) The distribution of the continuous gauge regions in the reduced BZ. Top: The BZ of a 2D hexagonal lattice with $C_{6v}$ symmetry, its reduced BZ is a triangle, and there are two situations for the distribution of the continuous gauge regions: Case 1 and Case 2 represent complete and non-complete continuous gauge regions within the reduced BZ respectively. Bottom: An ice cream cone-like zone, where two sides of the reduced BZ are connected, and its bottom is connected to its translational mirror part, forming a closed manifold, which is convenient for calculation. (b) The distribution of non-complete continuous gauge regions in the reduced BZ for the hexagonal and square lattice systems. The arcs in each picture represent the boundary between two independent continuous regions in the reduced BZs. Cases \uppercase\expandafter{\romannumeral1} to \uppercase\expandafter{\romannumeral4} in the left column represent all possible distributions of non-complete continuous gauge regions within the reduced BZ of a hexagonal lattice system. Cases \uppercase\expandafter{\romannumeral1} to \uppercase\expandafter{\romannumeral6} in the right part represent that of the square lattice system.
}
\end{figure}

After analyzing the far-field properties of the bulk and edge state,  the profound physics of topological phase transition should be revealed: after band crossing, the local exchange of the gauged region leads to the nontrivial topology.
To illustrate it, we need to know how symmetry constraint the Chern number, similar to the fact that symmetry has restrictions on BIC \cite{ref15}.
After calculating the Chern number using Kohmoto's method, it is easy to see that the distribution of continuous and fully gauged regions in the reduced BZ can be divided into two cases, as shown in Fig. 5(a): one with continuous and fully gauged regions in the whole reduced BZ(Case 1), and one with several parts of such regions exist (Case 2) (described in Supplementary Material Section  \Rmnum{5}). The Chern number $C$ of energy bands is determined by:
\begin{equation}
C=C_0+nm, 
\end{equation}
where $n$ represents the order of rotational symmetry for the crystal, $m$ is an integer determined by Case 1, $C_0$ is related to the representation carried by high-symmetry points, and the distribution of the incomplete and continuous gauged regions (Fig. 5(b)), which is determined by Case 2. 
After band inversion near the gamma point, the gauged region of case 1 will be exchanged, resulting in a change in the topological phase.

The above discussion is based on non-degenerate energy bands, where the Chern number can be well defined. However, in the case of degenerate bands, non-Abelian connections will emerge. By introducing the matrix form of the Wilson loop, the sum of the pumping numbers and winding numbers of all degenerate bands have a corresponding relationship (Supplementary Material Section \Rmnum{6}):          
\begin{equation}
\begin{array}{c}
\begin{aligned}
\sum_{n} P u m_{n} & =\oint d k \sum_{i, j, n} R_{j n} R_{n i}^{\dagger}\left\langle C_{i}(k)\left|i \nabla_{k}\right| C_{j}(k)\right\rangle+C \\
& =\sum_{i} q_{i}+C.
\end{aligned}
\end{array}
\end{equation}

\section{Conclusion}
In su, there are connections between the topological properties of energy bands and the far-field signals due to the mapping connections between different representation spaces of the same group representation. During adiabatic deformation, the winding and Chern numbers are topological invariants of the photonic crystal slab. It's also found that the symmetry of a system will influence the Chern numbers. In addition, the process of topological phase transition is depicted with the exchange of far-field polarization singularities for a system with pseudo-time-reversal symmetry. To further validate the robustness of the winding number, a three-layered photonic slab is presented, and its eigenmodes show that its winding number is fixed during adiabatic deformation. Finally, the edge states, one of the most concerned near-field effects in topology, are acquired by solving the Dirac equation, and the variation of quality factors for these edge states is studied. This work will inspire studies on the topological properties of far-field signals \cite{ref34,ref35,ref36,ref37,ref38} and extend the research scope from near-field to far-field effects for topological insulators, bringing potential applications for topological single-mode lasers \cite{ref39} and high-order harmonic generation of topological photon modes inside the light cone \cite{ref40,ref41}.

\section{Acknowledgements}
This work was supported by the National Natural Science Foundation of China (No. 12174052).

\bibliography{sn-bibliography}

\providecommand{\latin}[1]{#1}
\makeatletter
\providecommand{\doi}
  {\begingroup\let\do\@makeother\dospecials
  \catcode`\{=1 \catcode`\}=2 \doi@aux}
\providecommand{\doi@aux}[1]{\endgroup\texttt{#1}}
\makeatother
\providecommand*\mcitethebibliography{\thebibliography}
\csname @ifundefined\endcsname{endmcitethebibliography}
  {\let\endmcitethebibliography\endthebibliography}{}
\begin{mcitethebibliography}{50}
\providecommand*\natexlab[1]{#1}
\providecommand*\mciteSetBstSublistMode[1]{}
\providecommand*\mciteSetBstMaxWidthForm[2]{}
\providecommand*\mciteBstWouldAddEndPuncttrue
  {\def\EndOfBibitem{\unskip.}}
\providecommand*\mciteBstWouldAddEndPunctfalse
  {\let\EndOfBibitem\relax}
\providecommand*\mciteSetBstMidEndSepPunct[3]{}
\providecommand*\mciteSetBstSublistLabelBeginEnd[3]{}
\providecommand*\EndOfBibitem{}
\mciteSetBstSublistMode{f}
\mciteSetBstMaxWidthForm{subitem}{(\alph{mcitesubitemcount})}
\mciteSetBstSublistLabelBeginEnd
  {\mcitemaxwidthsubitemform\space}
  {\relax}
  {\relax}

\bibitem[v.~Klitzing \latin{et~al.}(1980)v.~Klitzing, Dorda, and Pepper]{ref1}
v.~Klitzing,~K.; Dorda,~G.; Pepper,~M. New {{Method}} for {{High-Accuracy
  Determination}} of the {{Fine-Structure Constant Based}} on {{Quantized Hall
  Resistance}}. \emph{Physical Review Letters} \textbf{1980}, \emph{45},
  494--497\relax
\mciteBstWouldAddEndPuncttrue
\mciteSetBstMidEndSepPunct{\mcitedefaultmidpunct}
{\mcitedefaultendpunct}{\mcitedefaultseppunct}\relax
\EndOfBibitem
\bibitem[Lan \latin{et~al.}(2022)Lan, Chen, Gao, Zhang, and Sha]{ref2}
Lan,~Z.; Chen,~M.~L.; Gao,~F.; Zhang,~S.; Sha,~W.~E. A brief review of
  topological photonics in one, two, and three dimensions. \emph{Reviews in
  Physics} \textbf{2022}, \emph{9}, 100076\relax
\mciteBstWouldAddEndPuncttrue
\mciteSetBstMidEndSepPunct{\mcitedefaultmidpunct}
{\mcitedefaultendpunct}{\mcitedefaultseppunct}\relax
\EndOfBibitem
\bibitem[Poli \latin{et~al.}(2015)Poli, Bellec, Kuhl, Mortessagne, and
  Schomerus]{ref3}
Poli,~C.; Bellec,~M.; Kuhl,~U.; Mortessagne,~F.; Schomerus,~H. Selective
  Enhancement of Topologically Induced Interface States in a Dielectric
  Resonator Chain. \emph{Nature Communications} \textbf{2015}, \emph{6},
  6710\relax
\mciteBstWouldAddEndPuncttrue
\mciteSetBstMidEndSepPunct{\mcitedefaultmidpunct}
{\mcitedefaultendpunct}{\mcitedefaultseppunct}\relax
\EndOfBibitem
\bibitem[Kiriushechkina \latin{et~al.}(2023)Kiriushechkina, Vakulenko,
  Smirnova, Guddala, Kawaguchi, Komissarenko, Allen, Allen, and
  Khanikaev]{ref4}
Kiriushechkina,~S.; Vakulenko,~A.; Smirnova,~D.; Guddala,~S.; Kawaguchi,~Y.;
  Komissarenko,~F.; Allen,~M.; Allen,~J.; Khanikaev,~A.~B. Spin-dependent
  properties of optical modes guided by adiabatic trapping potentials in
  photonic Dirac metasurfaces. \emph{Nature Nanotechnology} \textbf{2023},
  \emph{18}, 875--881\relax
\mciteBstWouldAddEndPuncttrue
\mciteSetBstMidEndSepPunct{\mcitedefaultmidpunct}
{\mcitedefaultendpunct}{\mcitedefaultseppunct}\relax
\EndOfBibitem
\bibitem[He \latin{et~al.}(2019)He, Liang, Yuan, Qiu, Chen, Zhao, and
  Dong]{ref5}
He,~X.-T.; Liang,~E.-T.; Yuan,~J.-J.; Qiu,~H.-Y.; Chen,~X.-D.; Zhao,~F.-L.;
  Dong,~J.-W. A silicon-on-insulator slab for topological valley transport.
  \emph{Nature Communications} \textbf{2019}, \emph{10}, 872\relax
\mciteBstWouldAddEndPuncttrue
\mciteSetBstMidEndSepPunct{\mcitedefaultmidpunct}
{\mcitedefaultendpunct}{\mcitedefaultseppunct}\relax
\EndOfBibitem
\bibitem[Xie \latin{et~al.}(2019)Xie, Su, Wang, Su, Shen, Zhan, Lu, Wang, and
  Chen]{ref6}
Xie,~B.-Y.; Su,~G.-X.; Wang,~H.-F.; Su,~H.; Shen,~X.-P.; Zhan,~P.; Lu,~M.-H.;
  Wang,~Z.-L.; Chen,~Y.-F. Visualization of higher-order topological insulating
  phases in two-dimensional dielectric photonic crystals. \emph{Physical Review
  Letters} \textbf{2019}, \emph{122}, 233903\relax
\mciteBstWouldAddEndPuncttrue
\mciteSetBstMidEndSepPunct{\mcitedefaultmidpunct}
{\mcitedefaultendpunct}{\mcitedefaultseppunct}\relax
\EndOfBibitem
\bibitem[Guo \latin{et~al.}(2017)Guo, Yang, Xia, Gao, Liu, Chen, Xiang, and
  Zhang]{ref7}
Guo,~Q.; Yang,~B.; Xia,~L.; Gao,~W.; Liu,~H.; Chen,~J.; Xiang,~Y.; Zhang,~S.
  Three dimensional photonic Dirac points in metamaterials. \emph{Physical
  Review Letters} \textbf{2017}, \emph{119}, 213901\relax
\mciteBstWouldAddEndPuncttrue
\mciteSetBstMidEndSepPunct{\mcitedefaultmidpunct}
{\mcitedefaultendpunct}{\mcitedefaultseppunct}\relax
\EndOfBibitem
\bibitem[Sun \latin{et~al.}(2017)Sun, Luo, Gong, Guo, and Zhou]{ref8}
Sun,~B.~Y.; Luo,~X.~W.; Gong,~M.; Guo,~G.~C.; Zhou,~Z.~W. Weyl semimetal phases
  and implementation in degenerate optical cavities. \emph{Physical Review A}
  \textbf{2017}, \emph{96}, 013857\relax
\mciteBstWouldAddEndPuncttrue
\mciteSetBstMidEndSepPunct{\mcitedefaultmidpunct}
{\mcitedefaultendpunct}{\mcitedefaultseppunct}\relax
\EndOfBibitem
\bibitem[Yan \latin{et~al.}(2018)Yan, Liu, Yan, Liu, Chen, Wang, and Lu]{ref9}
Yan,~Q.; Liu,~R.; Yan,~Z.; Liu,~B.; Chen,~H.; Wang,~Z.; Lu,~L. Experimental
  discovery of nodal chains. \emph{Nature Physics} \textbf{2018}, \emph{14},
  461--464\relax
\mciteBstWouldAddEndPuncttrue
\mciteSetBstMidEndSepPunct{\mcitedefaultmidpunct}
{\mcitedefaultendpunct}{\mcitedefaultseppunct}\relax
\EndOfBibitem
\bibitem[Liu \latin{et~al.}(2021)Liu, Yan, Xiao, and Zhu]{ref10}
Liu,~H.; Yan,~Z.; Xiao,~M.; Zhu,~S. Recent progress in synthetic dimension in
  topological photonics. \emph{Acta Optica Sinica} \textbf{2021}, \emph{41},
  0123002\relax
\mciteBstWouldAddEndPuncttrue
\mciteSetBstMidEndSepPunct{\mcitedefaultmidpunct}
{\mcitedefaultendpunct}{\mcitedefaultseppunct}\relax
\EndOfBibitem
\bibitem[Lu \latin{et~al.}(2014)Lu, Joannopoulos, and
  Solja{\v{c}}i{\'c}]{ref11}
Lu,~L.; Joannopoulos,~J.~D.; Solja{\v{c}}i{\'c},~M. Topological photonics.
  \emph{Nature photonics} \textbf{2014}, \emph{8}, 821--829\relax
\mciteBstWouldAddEndPuncttrue
\mciteSetBstMidEndSepPunct{\mcitedefaultmidpunct}
{\mcitedefaultendpunct}{\mcitedefaultseppunct}\relax
\EndOfBibitem
\bibitem[Gorlach \latin{et~al.}(2018)Gorlach, Ni, Smirnova, Korobkin, Zhirihin,
  Slobozhanyuk, Belov, Al{\`u}, and Khanikaev]{ref12}
Gorlach,~M.~A.; Ni,~X.; Smirnova,~D.~A.; Korobkin,~D.; Zhirihin,~D.;
  Slobozhanyuk,~A.~P.; Belov,~P.~A.; Al{\`u},~A.; Khanikaev,~A.~B. Far-field
  probing of leaky topological states in all-dielectric metasurfaces.
  \emph{Nature Communications} \textbf{2018}, \emph{9}, 909\relax
\mciteBstWouldAddEndPuncttrue
\mciteSetBstMidEndSepPunct{\mcitedefaultmidpunct}
{\mcitedefaultendpunct}{\mcitedefaultseppunct}\relax
\EndOfBibitem
\bibitem[F{\"o}sel \latin{et~al.}(2017)F{\"o}sel, Peano, and Marquardt]{ref13}
F{\"o}sel,~T.; Peano,~V.; Marquardt,~F. L lines, C points and Chern numbers:
  understanding band structure topology using polarization fields. \emph{New
  Journal of Physics} \textbf{2017}, \emph{19}, 115013\relax
\mciteBstWouldAddEndPuncttrue
\mciteSetBstMidEndSepPunct{\mcitedefaultmidpunct}
{\mcitedefaultendpunct}{\mcitedefaultseppunct}\relax
\EndOfBibitem
\bibitem[Yao \latin{et~al.}(2023)Yao, Li, Zhang, Li, Yue, Xu, and Yang]{ref14}
Yao,~J.-q.; Li,~J.-t.; Zhang,~Y.-t.; Li,~J.; Yue,~Z.; Xu,~H.; Yang,~F. Bound
  States in Continuum in Periodic Optical Systems. \emph{Chin. Opt}
  \textbf{2023}, \emph{16}, 1--23\relax
\mciteBstWouldAddEndPuncttrue
\mciteSetBstMidEndSepPunct{\mcitedefaultmidpunct}
{\mcitedefaultendpunct}{\mcitedefaultseppunct}\relax
\EndOfBibitem
\bibitem[Zhen \latin{et~al.}(2014)Zhen, Hsu, Lu, Stone, and
  Solja{\v{c}}i{\'c}]{ref15}
Zhen,~B.; Hsu,~C.~W.; Lu,~L.; Stone,~A.~D.; Solja{\v{c}}i{\'c},~M. Topological
  nature of optical bound states in the continuum. \emph{Physical Review
  Letters} \textbf{2014}, \emph{113}, 257401\relax
\mciteBstWouldAddEndPuncttrue
\mciteSetBstMidEndSepPunct{\mcitedefaultmidpunct}
{\mcitedefaultendpunct}{\mcitedefaultseppunct}\relax
\EndOfBibitem
\bibitem[Sun \latin{et~al.}(2024)Sun, Wang, Li, Cui, Chen, and Zhang]{ref16}
Sun,~G.; Wang,~Y.; Li,~Y.; Cui,~Z.; Chen,~W.; Zhang,~K. Tailoring topological
  nature of merging bound states in the continuum by manipulating structure
  symmetry of the all-dielectric metasurface. \emph{Physical Review B}
  \textbf{2024}, \emph{109}, 035406\relax
\mciteBstWouldAddEndPuncttrue
\mciteSetBstMidEndSepPunct{\mcitedefaultmidpunct}
{\mcitedefaultendpunct}{\mcitedefaultseppunct}\relax
\EndOfBibitem
\bibitem[Huang \latin{et~al.}(2024)Huang, Li, Zhou, Zhong, You, Li, Cheng, and
  Miroshnichenko]{ref17}
Huang,~L.; Li,~S.; Zhou,~C.; Zhong,~H.; You,~S.; Li,~L.; Cheng,~Y.;
  Miroshnichenko,~A.~E. Realizing Ultrahigh-Q Resonances Through Harnessing
  Symmetry-Protected Bound States in the Continuum. \emph{Advanced Functional
  Materials} \textbf{2024}, \emph{34}, 2309982\relax
\mciteBstWouldAddEndPuncttrue
\mciteSetBstMidEndSepPunct{\mcitedefaultmidpunct}
{\mcitedefaultendpunct}{\mcitedefaultseppunct}\relax
\EndOfBibitem
\bibitem[Liu \latin{et~al.}(2019)Liu, Wang, Zhang, Wang, Zhao, Guan, Liu, Shi,
  and Zi]{ref18}
Liu,~W.; Wang,~B.; Zhang,~Y.; Wang,~J.; Zhao,~M.; Guan,~F.; Liu,~X.; Shi,~L.;
  Zi,~J. Circularly polarized states spawning from bound states in the
  continuum. \emph{Physical Review Letters} \textbf{2019}, \emph{123},
  116104\relax
\mciteBstWouldAddEndPuncttrue
\mciteSetBstMidEndSepPunct{\mcitedefaultmidpunct}
{\mcitedefaultendpunct}{\mcitedefaultseppunct}\relax
\EndOfBibitem
\bibitem[Zhang \latin{et~al.}(2018)Zhang, Chen, Liu, Hsu, Wang, Guan, Liu, Shi,
  Lu, and Zi]{ref19}
Zhang,~Y.; Chen,~A.; Liu,~W.; Hsu,~C.~W.; Wang,~B.; Guan,~F.; Liu,~X.; Shi,~L.;
  Lu,~L.; Zi,~J. Observation of polarization vortices in momentum space.
  \emph{Physical Review Letters} \textbf{2018}, \emph{120}, 186103\relax
\mciteBstWouldAddEndPuncttrue
\mciteSetBstMidEndSepPunct{\mcitedefaultmidpunct}
{\mcitedefaultendpunct}{\mcitedefaultseppunct}\relax
\EndOfBibitem
\bibitem[Zeng \latin{et~al.}(2021)Zeng, Hu, Liu, Tang, and Qiu]{ref20}
Zeng,~Y.; Hu,~G.; Liu,~K.; Tang,~Z.; Qiu,~C.-W. Dynamics of topological
  polarization singularity in momentum space. \emph{Physical Review Letters}
  \textbf{2021}, \emph{127}, 176101\relax
\mciteBstWouldAddEndPuncttrue
\mciteSetBstMidEndSepPunct{\mcitedefaultmidpunct}
{\mcitedefaultendpunct}{\mcitedefaultseppunct}\relax
\EndOfBibitem
\bibitem[Yin \latin{et~al.}(2023)Yin, Inoue, Peng, and Noda]{ref21}
Yin,~X.; Inoue,~T.; Peng,~C.; Noda,~S. Topological unidirectional guided
  resonances emerged from interband coupling. \emph{Physical Review Letters}
  \textbf{2023}, \emph{130}, 056401\relax
\mciteBstWouldAddEndPuncttrue
\mciteSetBstMidEndSepPunct{\mcitedefaultmidpunct}
{\mcitedefaultendpunct}{\mcitedefaultseppunct}\relax
\EndOfBibitem
\bibitem[Zhang \latin{et~al.}(2021)Zhang, Lan, Xie, Chen, Sha, and Xu]{ref22}
Zhang,~Z.; Lan,~Z.; Xie,~Y.; Chen,~M.~L.; Sha,~W.~E.; Xu,~Y. Bound topological
  edge state in the continuum for all-dielectric photonic crystals.
  \emph{Physical Review Applied} \textbf{2021}, \emph{16}, 064036\relax
\mciteBstWouldAddEndPuncttrue
\mciteSetBstMidEndSepPunct{\mcitedefaultmidpunct}
{\mcitedefaultendpunct}{\mcitedefaultseppunct}\relax
\EndOfBibitem
\bibitem[Hu \latin{et~al.}(2021)Hu, Zhang, Haixiao, Zheng, Xiong, Yue, Wang,
  Xu, Cheng, Liu, and Christensen]{ref23}
Hu,~B.; Zhang,~Z.; Haixiao,~Z.; Zheng,~L.-Y.; Xiong,~W.; Yue,~Z.; Wang,~X.;
  Xu,~J.; Cheng,~Y.; Liu,~X.; Christensen,~J. Non-Hermitian topological
  whispering gallery. \emph{Nature} \textbf{2021}, \emph{597}, 655--659\relax
\mciteBstWouldAddEndPuncttrue
\mciteSetBstMidEndSepPunct{\mcitedefaultmidpunct}
{\mcitedefaultendpunct}{\mcitedefaultseppunct}\relax
\EndOfBibitem
\bibitem[Li \latin{et~al.}(2023)Li, Wei, Cotrufo, Chen, Mann, Ni, Xu, Chen,
  Wang, Fan, \latin{et~al.} others]{ref24}
Li,~A.; Wei,~H.; Cotrufo,~M.; Chen,~W.; Mann,~S.; Ni,~X.; Xu,~B.; Chen,~J.;
  Wang,~J.; Fan,~S.; others Exceptional Points and Non-{{Hermitian}} Photonics
  at the Nanoscale. \emph{Nature Nanotechnology} \textbf{2023}, \emph{18},
  706--720\relax
\mciteBstWouldAddEndPuncttrue
\mciteSetBstMidEndSepPunct{\mcitedefaultmidpunct}
{\mcitedefaultendpunct}{\mcitedefaultseppunct}\relax
\EndOfBibitem
\bibitem[Xue \latin{et~al.}(2022)Xue, Yang, and Zhang]{ref25}
Xue,~H.; Yang,~Y.; Zhang,~B. Topological acoustics. \emph{Nature Reviews
  Materials} \textbf{2022}, \emph{7}, 974--990\relax
\mciteBstWouldAddEndPuncttrue
\mciteSetBstMidEndSepPunct{\mcitedefaultmidpunct}
{\mcitedefaultendpunct}{\mcitedefaultseppunct}\relax
\EndOfBibitem
\bibitem[Chen \latin{et~al.}(2016)Chen, Taylor, and Yu]{ref26}
Chen,~H.-T.; Taylor,~A.~J.; Yu,~N. A review of metasurfaces: physics and
  applications. \emph{Reports on Progress in Physics} \textbf{2016}, \emph{79},
  076401\relax
\mciteBstWouldAddEndPuncttrue
\mciteSetBstMidEndSepPunct{\mcitedefaultmidpunct}
{\mcitedefaultendpunct}{\mcitedefaultseppunct}\relax
\EndOfBibitem
\bibitem[Zhang \latin{et~al.}(2024)Zhang, Shi, Liu, Tang, Zhang, and
  Dong]{ref27}
Zhang,~J.; Shi,~W.; Liu,~A.; Tang,~L.; Zhang,~S.; Dong,~Z. Amplitude and Phase
  Modulation with Electric Quadrupole Radiation. \emph{Journal of Applied
  Physics} \textbf{2024}, \emph{135}, 143104\relax
\mciteBstWouldAddEndPuncttrue
\mciteSetBstMidEndSepPunct{\mcitedefaultmidpunct}
{\mcitedefaultendpunct}{\mcitedefaultseppunct}\relax
\EndOfBibitem
\bibitem[Wu and Hu(2015)Wu, and Hu]{ref28}
Wu,~L.-H.; Hu,~X. Scheme for achieving a topological photonic crystal by using
  dielectric material. \emph{Physical Review Letters} \textbf{2015},
  \emph{114}, 223901\relax
\mciteBstWouldAddEndPuncttrue
\mciteSetBstMidEndSepPunct{\mcitedefaultmidpunct}
{\mcitedefaultendpunct}{\mcitedefaultseppunct}\relax
\EndOfBibitem
\bibitem[Zhu \latin{et~al.}(2018)Zhu, Wang, Xu, Lai, Jiang, and John]{ref29}
Zhu,~X.; Wang,~H.-X.; Xu,~C.; Lai,~Y.; Jiang,~J.-H.; John,~S. Topological
  transitions in continuously deformed photonic crystals. \emph{Physical Review
  B} \textbf{2018}, \emph{97}, 085148\relax
\mciteBstWouldAddEndPuncttrue
\mciteSetBstMidEndSepPunct{\mcitedefaultmidpunct}
{\mcitedefaultendpunct}{\mcitedefaultseppunct}\relax
\EndOfBibitem
\bibitem[Joannopoulos \latin{et~al.}(2011)Joannopoulos, Johnson, Winn, and
  Meade]{ref30}
Joannopoulos,~J.~D.; Johnson,~S.~G.; Winn,~J.~N.; Meade,~R.~D. Photonic
  {{Crystals}}: {{Molding}} the {{Flow}} of {{Light}} - {{Second Edition}}.
  2011\relax
\mciteBstWouldAddEndPuncttrue
\mciteSetBstMidEndSepPunct{\mcitedefaultmidpunct}
{\mcitedefaultendpunct}{\mcitedefaultseppunct}\relax
\EndOfBibitem
\bibitem[Kang \latin{et~al.}(2022)Kang, Mao, Zhang, Xiao, Xu, and Chan]{ref31}
Kang,~M.; Mao,~L.; Zhang,~S.; Xiao,~M.; Xu,~H.; Chan,~C.~T. Merging bound
  states in the continuum by harnessing higher-order topological charges.
  \emph{Light: Science \& Applications} \textbf{2022}, \emph{11}, 228\relax
\mciteBstWouldAddEndPuncttrue
\mciteSetBstMidEndSepPunct{\mcitedefaultmidpunct}
{\mcitedefaultendpunct}{\mcitedefaultseppunct}\relax
\EndOfBibitem
\bibitem[Zhao \latin{et~al.}(2020)Zhao, Xie, Chen, Lan, Huang, and Sha]{ref32}
Zhao,~R.; Xie,~G.-D.; Chen,~M. L.~N.; Lan,~Z.; Huang,~Z.; Sha,~W. E.~I.
  First-Principle Calculation of {{Chern}} Number in Gyrotropic Photonic
  Crystals. \emph{Optics Express} \textbf{2020}, \emph{28}, 4638--4649\relax
\mciteBstWouldAddEndPuncttrue
\mciteSetBstMidEndSepPunct{\mcitedefaultmidpunct}
{\mcitedefaultendpunct}{\mcitedefaultseppunct}\relax
\EndOfBibitem
\bibitem[Murakami(2007)]{murakami2007phase}
Murakami,~S. Phase Transition between the Quantum Spin {{Hall}} and Insulator
  Phases in {{3D}}: Emergence of a Topological Gapless Phase. \emph{New Journal
  of Physics} \textbf{2007}, \emph{9}, 356\relax
\mciteBstWouldAddEndPuncttrue
\mciteSetBstMidEndSepPunct{\mcitedefaultmidpunct}
{\mcitedefaultendpunct}{\mcitedefaultseppunct}\relax
\EndOfBibitem
\bibitem[Bulgakov and Maksimov(2017)Bulgakov, and Maksimov]{ref34}
Bulgakov,~E.~N.; Maksimov,~D.~N. Topological bound states in the continuum in
  arrays of dielectric spheres. \emph{Physical Review Letters} \textbf{2017},
  \emph{118}, 267401\relax
\mciteBstWouldAddEndPuncttrue
\mciteSetBstMidEndSepPunct{\mcitedefaultmidpunct}
{\mcitedefaultendpunct}{\mcitedefaultseppunct}\relax
\EndOfBibitem
\bibitem[Zhao \latin{et~al.}(2022)Zhao, Dong, Zhang, Zeng, Hu, and
  Zhang]{ref35}
Zhao,~C.; Dong,~S.; Zhang,~Q.; Zeng,~Y.; Hu,~G.; Zhang,~Y. Magnetic modulation
  of topological polarization singularities in momentum space. \emph{Optics
  Letters} \textbf{2022}, \emph{47}, 2754--2757\relax
\mciteBstWouldAddEndPuncttrue
\mciteSetBstMidEndSepPunct{\mcitedefaultmidpunct}
{\mcitedefaultendpunct}{\mcitedefaultseppunct}\relax
\EndOfBibitem
\bibitem[Weimann \latin{et~al.}(2013)Weimann, Xu, Keil, Miroshnichenko,
  T{\"u}nnermann, Nolte, Sukhorukov, Szameit, and Kivshar]{ref36}
Weimann,~S.; Xu,~Y.; Keil,~R.; Miroshnichenko,~A.~E.; T{\"u}nnermann,~A.;
  Nolte,~S.; Sukhorukov,~A.~A.; Szameit,~A.; Kivshar,~Y.~S. Compact surface
  Fano states embedded in the continuum of waveguide arrays. \emph{Physical
  Review Letters} \textbf{2013}, \emph{111}, 240403\relax
\mciteBstWouldAddEndPuncttrue
\mciteSetBstMidEndSepPunct{\mcitedefaultmidpunct}
{\mcitedefaultendpunct}{\mcitedefaultseppunct}\relax
\EndOfBibitem
\bibitem[Guo \latin{et~al.}(2020)Guo, Xiao, Guo, Yuan, and Fan]{ref37}
Guo,~C.; Xiao,~M.; Guo,~Y.; Yuan,~L.; Fan,~S. Meron spin textures in momentum
  space. \emph{Physical Review Letters} \textbf{2020}, \emph{124}, 106103\relax
\mciteBstWouldAddEndPuncttrue
\mciteSetBstMidEndSepPunct{\mcitedefaultmidpunct}
{\mcitedefaultendpunct}{\mcitedefaultseppunct}\relax
\EndOfBibitem
\bibitem[Chen \latin{et~al.}(2021)Chen, Yang, Chen, and Liu]{ref38}
Chen,~W.; Yang,~Q.; Chen,~Y.; Liu,~W. Evolution and global charge conservation
  for polarization singularities emerging from non-Hermitian degeneracies.
  \emph{Proceedings of the National Academy of Sciences} \textbf{2021},
  \emph{118}, e2019578118\relax
\mciteBstWouldAddEndPuncttrue
\mciteSetBstMidEndSepPunct{\mcitedefaultmidpunct}
{\mcitedefaultendpunct}{\mcitedefaultseppunct}\relax
\EndOfBibitem
\bibitem[Bandres \latin{et~al.}(2018)Bandres, Wittek, Harari, Parto, Ren,
  Segev, Christodoulides, and Khajavikhan]{ref39}
Bandres,~M.~A.; Wittek,~S.; Harari,~G.; Parto,~M.; Ren,~J.; Segev,~M.;
  Christodoulides,~D.~N.; Khajavikhan,~M. Topological insulator laser:
  Experiments. \emph{Science} \textbf{2018}, \emph{359}, eaar4005\relax
\mciteBstWouldAddEndPuncttrue
\mciteSetBstMidEndSepPunct{\mcitedefaultmidpunct}
{\mcitedefaultendpunct}{\mcitedefaultseppunct}\relax
\EndOfBibitem
\bibitem[Wang \latin{et~al.}(2020)Wang, Clementi, Minkov, Barone, Carlin,
  Grandjean, Gerace, Fan, Galli, and Houdr{\'e}]{ref40}
Wang,~J.; Clementi,~M.; Minkov,~M.; Barone,~A.; Carlin,~J.-F.; Grandjean,~N.;
  Gerace,~D.; Fan,~S.; Galli,~M.; Houdr{\'e},~R. Doubly resonant
  second-harmonic generation of a vortex beam from a bound state in the
  continuum. \emph{Optica} \textbf{2020}, \emph{7}, 1126--1132\relax
\mciteBstWouldAddEndPuncttrue
\mciteSetBstMidEndSepPunct{\mcitedefaultmidpunct}
{\mcitedefaultendpunct}{\mcitedefaultseppunct}\relax
\EndOfBibitem
\bibitem[Kruk \latin{et~al.}(2021)Kruk, Gao, Choi, Zentgraf, Zhang, and
  Kivshar]{ref41}
Kruk,~S.~S.; Gao,~W.; Choi,~D.-Y.; Zentgraf,~T.; Zhang,~S.; Kivshar,~Y.
  Nonlinear imaging of nanoscale topological corner states. \emph{Nano Letters}
  \textbf{2021}, \emph{21}, 4592--4597\relax
\mciteBstWouldAddEndPuncttrue
\mciteSetBstMidEndSepPunct{\mcitedefaultmidpunct}
{\mcitedefaultendpunct}{\mcitedefaultseppunct}\relax
\EndOfBibitem
\bibitem[Zhen \latin{et~al.}(2014)Zhen, Hsu, Lu, Stone, and
  Solja{\v{c}}i{\'c}]{r1}
Zhen,~B.; Hsu,~C.~W.; Lu,~L.; Stone,~A.~D.; Solja{\v{c}}i{\'c},~M. Topological
  nature of optical bound states in the continuum. \emph{Physical Review
  Letters} \textbf{2014}, \emph{113}, 257401\relax
\mciteBstWouldAddEndPuncttrue
\mciteSetBstMidEndSepPunct{\mcitedefaultmidpunct}
{\mcitedefaultendpunct}{\mcitedefaultseppunct}\relax
\EndOfBibitem
\bibitem[Shen(2017)]{r2}
Shen,~S.-Q. \emph{Topological {{Insulators}}: {{Dirac Equation}} in {{Condensed
  Matter}}}; Springer {{Series}} in {{Solid-State Sciences}}; Springer
  Singapore: Singapore, 2017; Vol. 187\relax
\mciteBstWouldAddEndPuncttrue
\mciteSetBstMidEndSepPunct{\mcitedefaultmidpunct}
{\mcitedefaultendpunct}{\mcitedefaultseppunct}\relax
\EndOfBibitem
\bibitem[Zhao \latin{et~al.}(2020)Zhao, Xie, Chen, Lan, Huang, and Sha]{r3}
Zhao,~R.; Xie,~G.-D.; Chen,~M. L.~N.; Lan,~Z.; Huang,~Z.; Sha,~W. E.~I.
  First-Principle Calculation of {{Chern}} Number in Gyrotropic Photonic
  Crystals. \emph{Optics Express} \textbf{2020}, \emph{28}, 4638--4649\relax
\mciteBstWouldAddEndPuncttrue
\mciteSetBstMidEndSepPunct{\mcitedefaultmidpunct}
{\mcitedefaultendpunct}{\mcitedefaultseppunct}\relax
\EndOfBibitem
\bibitem[Zhen \latin{et~al.}(2014)Zhen, Hsu, Lu, Stone, and
  Solja{\v{c}}i{\'c}]{rh}
Zhen,~B.; Hsu,~C.~W.; Lu,~L.; Stone,~A.~D.; Solja{\v{c}}i{\'c},~M. Topological
  nature of optical bound states in the continuum. \emph{Physical Review
  Letters} \textbf{2014}, \emph{113}, 257401\relax
\mciteBstWouldAddEndPuncttrue
\mciteSetBstMidEndSepPunct{\mcitedefaultmidpunct}
{\mcitedefaultendpunct}{\mcitedefaultseppunct}\relax
\EndOfBibitem
\bibitem[Joannopoulos \latin{et~al.}(2011)Joannopoulos, Johnson, Winn, and
  Meade]{r6}
Joannopoulos,~J.~D.; Johnson,~S.~G.; Winn,~J.~N.; Meade,~R.~D. Photonic
  {{Crystals}}: {{Molding}} the {{Flow}} of {{Light}} - {{Second Edition}}.
  2011\relax
\mciteBstWouldAddEndPuncttrue
\mciteSetBstMidEndSepPunct{\mcitedefaultmidpunct}
{\mcitedefaultendpunct}{\mcitedefaultseppunct}\relax
\EndOfBibitem
\bibitem[Wu and Hu(2015)Wu, and Hu]{r7}
Wu,~L.-H.; Hu,~X. Scheme for achieving a topological photonic crystal by using
  dielectric material. \emph{Physical Review Letters} \textbf{2015},
  \emph{114}, 223901\relax
\mciteBstWouldAddEndPuncttrue
\mciteSetBstMidEndSepPunct{\mcitedefaultmidpunct}
{\mcitedefaultendpunct}{\mcitedefaultseppunct}\relax
\EndOfBibitem
\bibitem[Kang \latin{et~al.}(2022)Kang, Mao, Zhang, Xiao, Xu, and Chan]{r4}
Kang,~M.; Mao,~L.; Zhang,~S.; Xiao,~M.; Xu,~H.; Chan,~C.~T. Merging bound
  states in the continuum by harnessing higher-order topological charges.
  \emph{Light: Science \& Applications} \textbf{2022}, \emph{11}, 228\relax
\mciteBstWouldAddEndPuncttrue
\mciteSetBstMidEndSepPunct{\mcitedefaultmidpunct}
{\mcitedefaultendpunct}{\mcitedefaultseppunct}\relax
\EndOfBibitem
\bibitem[Asb{\'o}th \latin{et~al.}(2016)Asb{\'o}th, Oroszl{\'a}ny, and
  P{\'a}lyi]{r5}
Asb{\'o}th,~J.~K.; Oroszl{\'a}ny,~L.; P{\'a}lyi,~A. \emph{A {{Short Course}} on
  {{Topological Insulators}}}; Lecture {{Notes}} in {{Physics}}; Springer
  International Publishing: Cham, 2016; Vol. 919\relax
\mciteBstWouldAddEndPuncttrue
\mciteSetBstMidEndSepPunct{\mcitedefaultmidpunct}
{\mcitedefaultendpunct}{\mcitedefaultseppunct}\relax
\EndOfBibitem
\end{mcitethebibliography}

\renewcommand{\theequation}{S\arabic{equation}} 
\renewcommand{\thefigure}{S\arabic{figure}}
\renewcommand{\thetable}{S\arabic{table}} 
latex
\newpage
\appendix
\setcounter{figure}{0}
\captionsetup[figure]{labelfont={bf},name={Fig.},labelsep=period}
\begin{singlespacing}
{\centering \Large \section{Supplementary Information for Topology of Far-field Signals for Photonic Crystal Slabs}}

\subsection{Correspondence between Far-field Polarization and Chern Numbers}
For photonic crystal slabs, the eigenstates within the light cone will radiate field to the remote region due to the lack of constraints in the vertical direction\nocite{r1}. In two-dimensional photonic crystal slabs, quasi-TE and quasi-TM modes can be defined, and this discussion primarily focuses on the TE mode. A correspondence exists between the far-field polarization state and the eigenstates of the photonic crystal slab, which could be used to calculate the Berry curvature \nocite{r2,r3}. Naturally, one would expect that there is a connection between The Chern number and far-field polarization. However, finding their relationship by solving Maxwell's equations is challenging, while here, we utilized the symmetries inherent in the system to explore their correspondence successfully. Based on group theory, the eigenstates $\phi _{n}(r)$ of the Hamiltonian of photonic crystals carry a representation of the symmetry group(using the Einstein summation rule):
\begin{equation}
\phi _{n}\left( \Lambda^{-1}r\right)=\hat{R} \left( \Lambda\right) _{nm}\phi _{m}\left( r\right).
\end{equation}

The far-field polarization state in the system, as a vector, will also follow the transformations below\nocite{rh}:
\begin{equation}
\Lambda^u{}_vC_{un}(k)=\hat{\Tilde{R}}(\Lambda)_{nm}C_{vm}(\Lambda k).
\end{equation}

For mathematical convenience, we introduce the complex number:
\begin{equation}
\vec{C}(k)=C_x\vec{x}+C_y\vec{y}\xrightarrow{}\vec{C}(k)=C_x\vec{x}+iC_y\vec{y}.
\end{equation}

The polarization direction rotation corresponds to a phase variation for ${C}(k)$. Like the equation S2, the far-field polarization state follows the transformation below, where $\Lambda_{1}$ or $\Lambda_{2}$ is a certain element in the symmetry group:
\begin{equation}
\Lambda_{1}{ }^{u}{ }_{v} C_{u n}(k)=\delta^{u}{ }_{v} e^{i \theta\left(n, m, k, \Lambda_{1}\right)} C_{u n}(k)=\hat{\widetilde{R}}\left(\Lambda_{1}\right)_{n m} C_{v m}\left(\Lambda_{1} k\right),
\end{equation}
\begin{equation}
\Lambda_{2}{ }^{u}{ }_{v} C_{u n}(k)=\delta^{u}{ }_{v} e^{i \theta\left(n, m, k, \Lambda_{2}\right)} C_{u n}(k)=\hat{\widetilde{R}}\left(\Lambda_{2}\right)_{n m} C_{v m}\left(\Lambda_{2} k\right),
\end{equation}
where $\theta\left(n, m, k, \Lambda\right)$originates from the rotation of $C(k)$, and it is a function related to $n,m,k,\Lambda$, with the following property:

\begin{equation}
\theta(n,m,k,\Lambda_1)+\theta(n,m,k,\Lambda_2)=\theta(n,m,k,\Lambda_1,\Lambda_2).
\end{equation}

According to the property of group elements, 

\begin{equation}
\Lambda_{1}{ }^{u}{ }_{v^{\prime}} \Lambda_{2}{ }^{v^{\prime}}{ }_{v} C_{u n}(k)=\hat{\tilde{R}}\left(\Lambda_{1}\right)_{n m^{\prime}} \hat{\tilde{R}}\left(\Lambda_{2}\right)_{m^{\prime} m} C_{v m}\left(\Lambda_{1} \Lambda_{2} k\right).
\end{equation}

The above discussion turns the vector field into a complex scalar field, assuming that $\Lambda_{1} \Lambda_{2}=\Lambda_{3}$ and $\hat{\tilde{R^{\prime}}}(\Lambda)_{n m}=\hat{\tilde{R}}(\Lambda)_{n m} e^{-i \theta(n, m, k, \Lambda)}$, and taking the results of equation (S4,S5) into it: $C_{n}(k)=\hat{\tilde{R}}^{\prime}\left(\Lambda_{3}\right)_{n m} C_{m}\left(\Lambda_{3} k\right)$, $ C_{n}(k)=\hat{\tilde{R}}^{\prime}\left(\Lambda_{1}\right)_{n m}  \hat{\tilde{R}}^{\prime}\left(\Lambda_{2}\right)_{m^{\prime} m} C_{m}\left(\Lambda_{1} \Lambda_{2} k\right)$
Which could be written in the following form:
\begin{equation}
C_{n}\left(\Lambda_{3}{ }^{-1} k\right)=\hat{\tilde{R}}^{\prime}\left(\Lambda_{3}\right)_{n m} C_{m}(k),
\end{equation}
\begin{equation}
C_n((\Lambda_1\Lambda_2)^{-1}k)=\hat{\Tilde{R'}}(\Lambda_1)_{nm'}\hat{\Tilde{R'}}(\Lambda_2)_{m'm}C_m(k),
\end{equation}
from which, the relation that  $\hat{\tilde{R}}^{\prime}\left(\Lambda_{1}\right) \hat{\tilde{R}^{\prime}}\left(\Lambda_{2}\right)=\hat{\tilde{R}}^{\prime}\left(\Lambda_{3}\right)$ is acquired, indicating that $\hat{\tilde{R}}^{\prime}\left(\Lambda_{1}\right)$ is a representation of a symmetry group, and comparing this relation to that of the eigenmodes of the Hamiltonian: $\phi_{n}\left(\Lambda^{-1} r\right)=\hat{R}(\Lambda)_{n m} \phi_{m}(r) ; C_{n}\left(\Lambda^{-1} k\right)=\hat{\tilde{R}}^{\prime}(\Lambda)_{n m} C_{m}(k)$, It can be observed that the eigenstates of the Hamiltonian $\phi(r)$, and $C(k)$ carry the same group representation, albeit in different linear spaces, and a one-to-one mapping between them can be obtained. For convenience, we define the normalized polarization state with the Dirac notation: $\langle C(k) \mid C(k)\rangle=C^{*}(k) C(k)=C_{x}^{2}+C_{y}^{2}=1$.
The mapping operation could be represented with a similarity transformation matrix formally:
\begin{equation}
|\phi\rangle_n=\sum_m\Gamma_{nm}(k)|C\rangle_m
\end{equation}
Based on this relation, the Chern number, which is the integral of the Berry connection of the eigenstates over the BZ boundary, could correspond to the integration of the far-field polarization angle variation around the boundary of the BZ: the phase shift between $|C(k+\Delta k)\rangle$ and  $|C(k)\rangle$ could be defined as $e^{-i \Delta}=\langle C(k+\Delta k) \mid C(k)\rangle$, therefore, $1-i \Delta=\left\langle C(k)\left|\nabla_{k}\right| C(k)\right\rangle d k+1$ , and the phase shift $\Delta$ could be written in the form of connection:
\begin{equation}
\triangle=i\langle C(k)|\nabla_k|C(k)\rangle dk,
\end{equation}
which is similar to the Berry connection:
\begin{equation}
A(k)dk=i\langle\phi(k)|\nabla_k|\phi(k)\rangle dk.
\end{equation}

For energy bands within the light cone, we can obtain the sum of topological charges for the far-field polarization in the BZ (winding number), including the C-points and BIC:

\begin{equation}
q=\oint_{BZ}i\langle C(k)|\nabla_k|C(k)\rangle dk/2\pi.
\end{equation}

For photonic crystal slab with $C_{nv}$ symmetry, the Chern number of TE modes under nondegenerate situations is
\begin{equation}
\begin{array}{c}
C=\oint_{B Z} i\left\langle\phi(k)\left|\nabla_{k}\right| \phi(k)\right\rangle d k / 2 \pi=\oint_{B Z} i\left\langle C(k)\left|\Gamma^{*}(k) \nabla_{k} \Gamma(k)\right| C(k)\right\rangle d k / 2 \pi \\
 \qquad=\oint_{B Z} i\left\langle C(k)\left|\Gamma^{*}(k)\left(\nabla_{k} \Gamma(k)\right)\right| C(k)\right\rangle d k / 2 \pi+\oint_{B Z} i\left\langle C(k)\left|\nabla_{k}\right| C(k)\right\rangle d k / 2 \pi \\
  =\oint_{B Z} i \Gamma^{*}(k)\left(\nabla_{k} \Gamma(k)\right) d k / 2 \pi+\oint_{B Z} i\left\langle C(k)\left|\nabla_{k}\right| C(k)\right\rangle d k / 2 \pi,
\end{array}
\end{equation}
where $\Gamma(k)$ is a single-valued function of $k$, and the first integration of the final step must be an integer. The second term corresponds to the winding number in equation S13, which is the sum of topological singularity charges: $q=\sum_{k} q_{k}$ . So, symmetry-protected BICs will not disappear or create during adiabatic deformation, where the energy gap remains open.

	\begin{figure}[htbp]
	\centering
	\begin{subfigure}[t]{0.33\textwidth}
		\captionsetup{justification=raggedright, singlelinecheck=false, labelformat=empty, skip=0pt, position=top}
		\caption*{(a)}
\includegraphics[height=3.2cm]{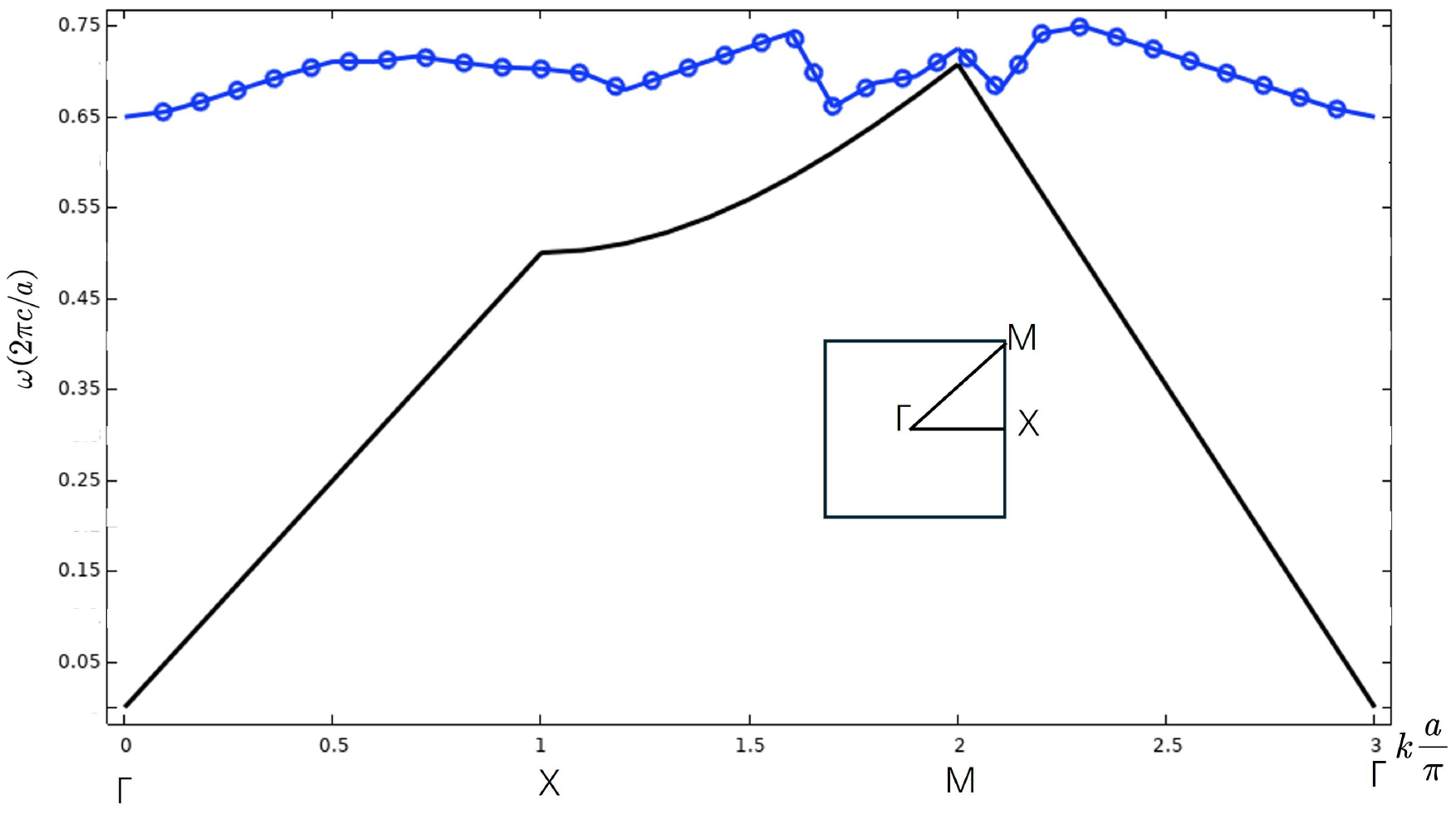}
		\label{fig:sub1}
	\end{subfigure}
	\begin{subfigure}[t]{0.24\textwidth}
		\captionsetup{justification=raggedright, singlelinecheck=false, labelformat=empty, skip=0pt, position=top}
		\caption*{(b)}
\includegraphics[height=3.2cm]{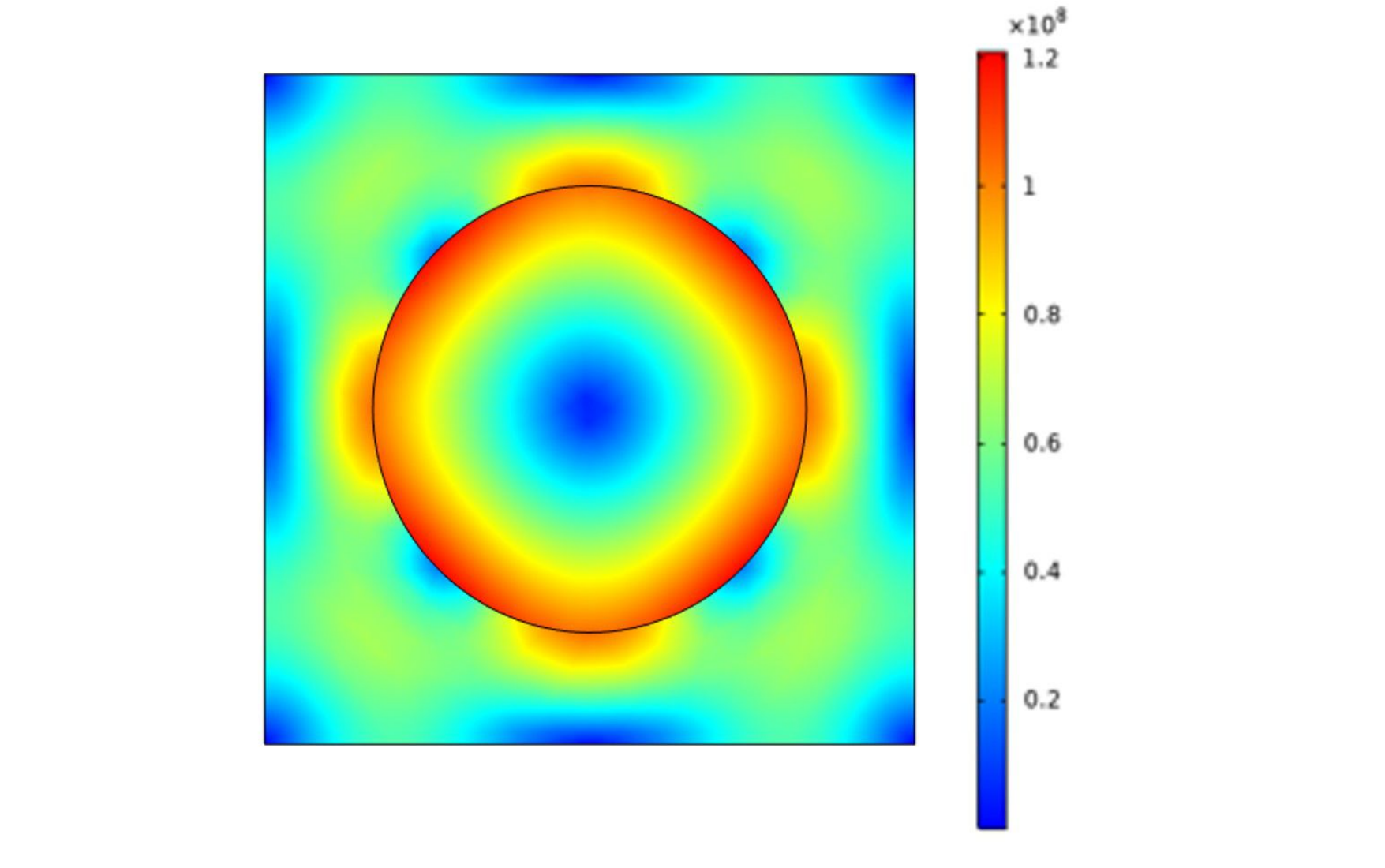}
		\label{fig:sub2}
	
	\end{subfigure}
	\begin{subfigure}[t]{0.4\textwidth}
		\captionsetup{justification=raggedright, singlelinecheck=false, labelformat=empty, skip=0pt, position=top}
		\caption*{(c)}
\includegraphics[height=3.2cm]{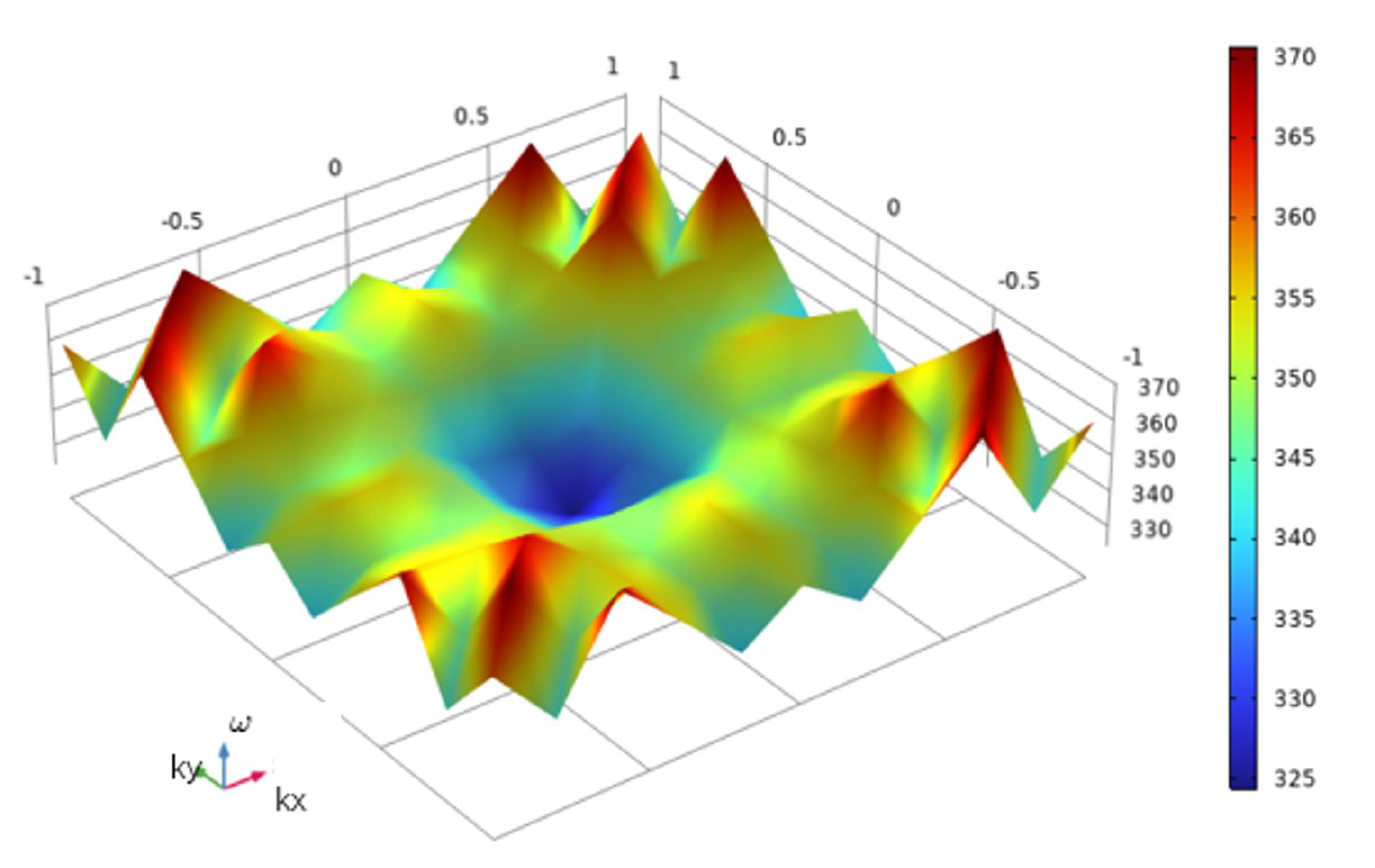}
		\label{fig:sub1}
	\end{subfigure}
\begin{subfigure}[t]{0.24\textwidth}
		\captionsetup{justification=raggedright, singlelinecheck=false, labelformat=empty, skip=0pt, position=top}
		\caption*{(d)}
\includegraphics[height=3.2cm]{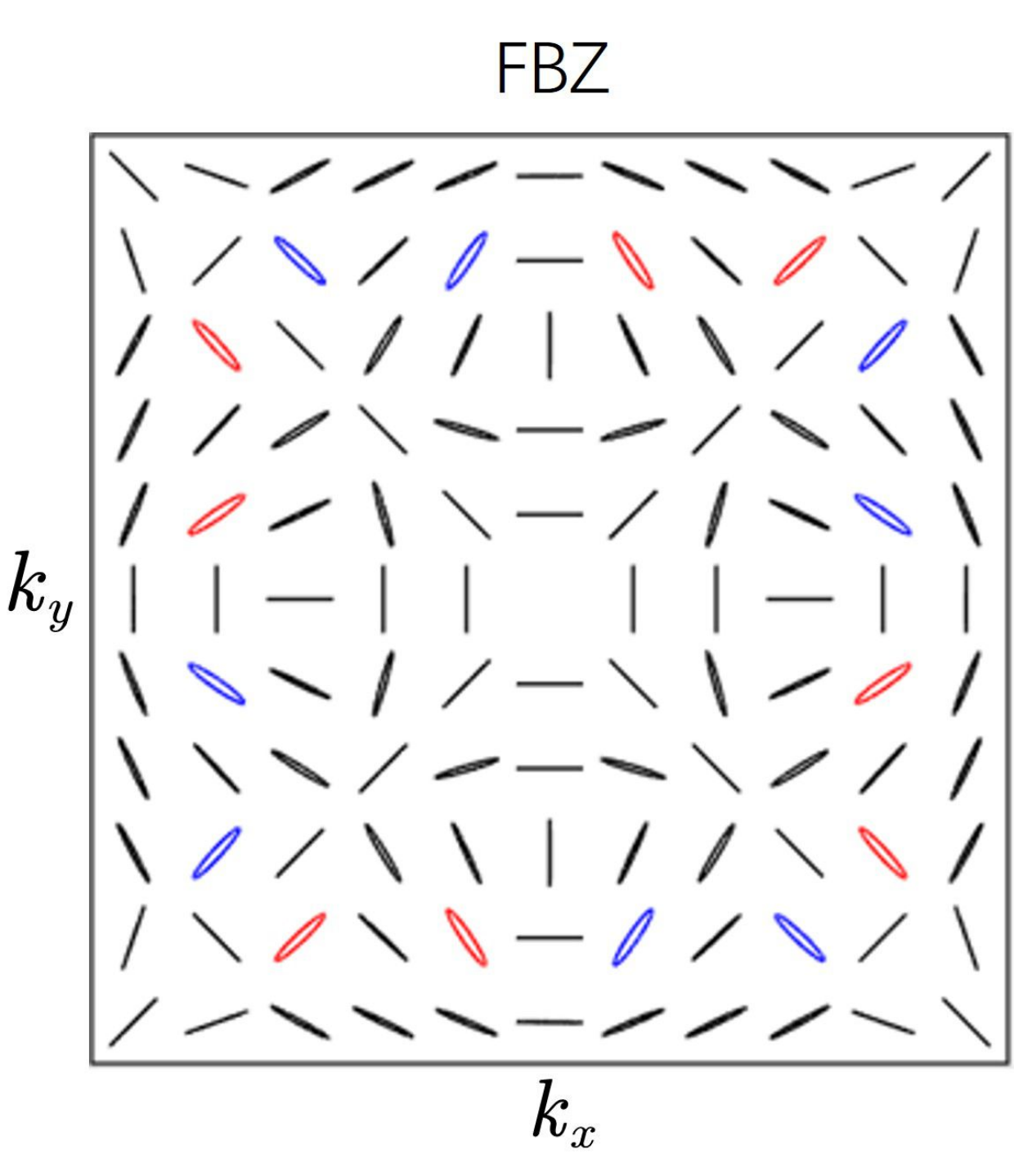}
		\label{fig:sub2}
	\end{subfigure}
		\begin{subfigure}[t]{0.4\textwidth}
		\captionsetup{justification=raggedright, singlelinecheck=false, labelformat=empty, skip=0pt, position=top}
		\caption*{(e)}
\includegraphics[height=3.2cm]{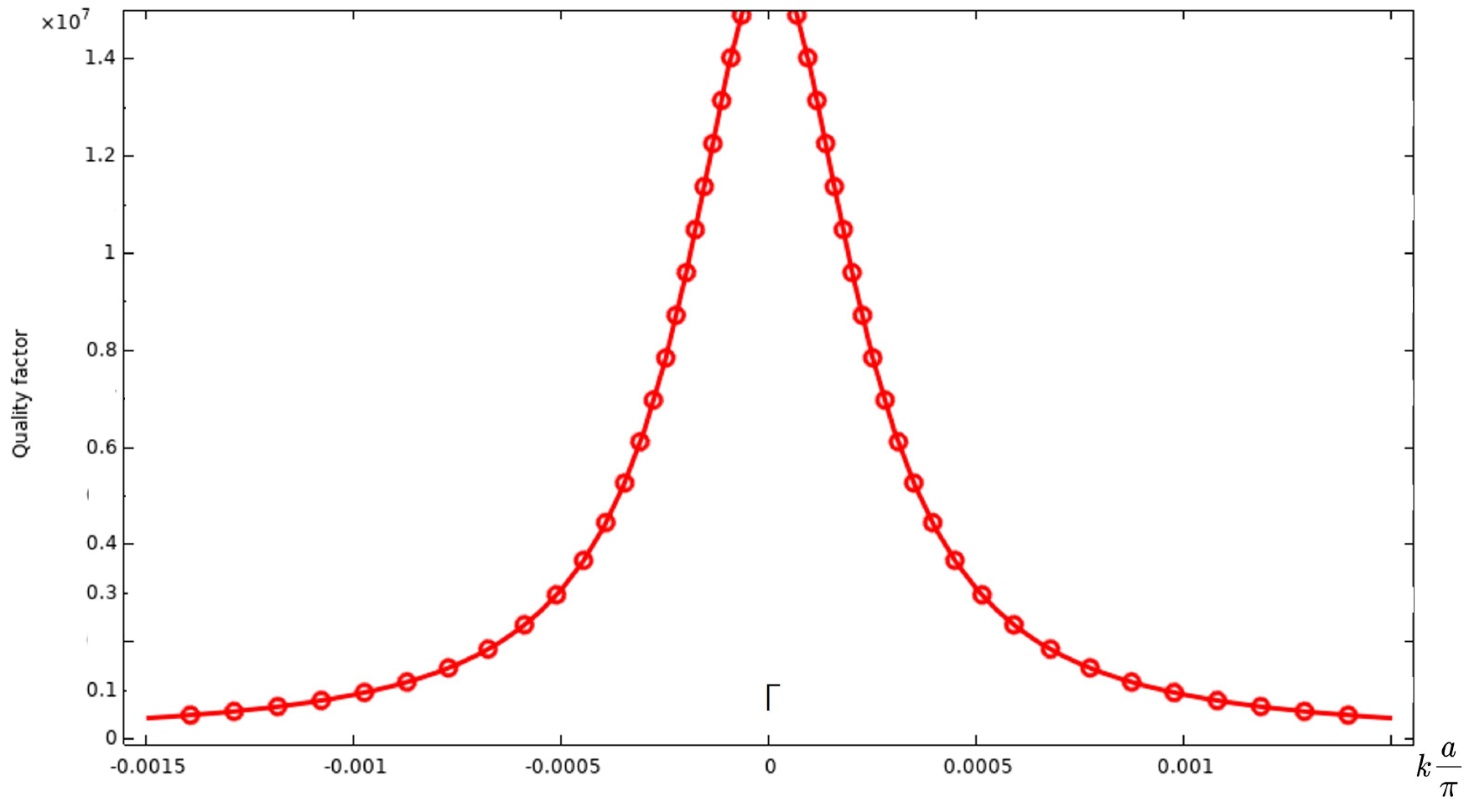}
		\label{fig:sub1}
	\end{subfigure}
	\begin{subfigure}[t]{0.24\textwidth}
		\captionsetup{justification=raggedright, singlelinecheck=false, labelformat=empty, skip=0pt, position=top}
		\caption*{(f)}
\includegraphics[height=3.2cm]{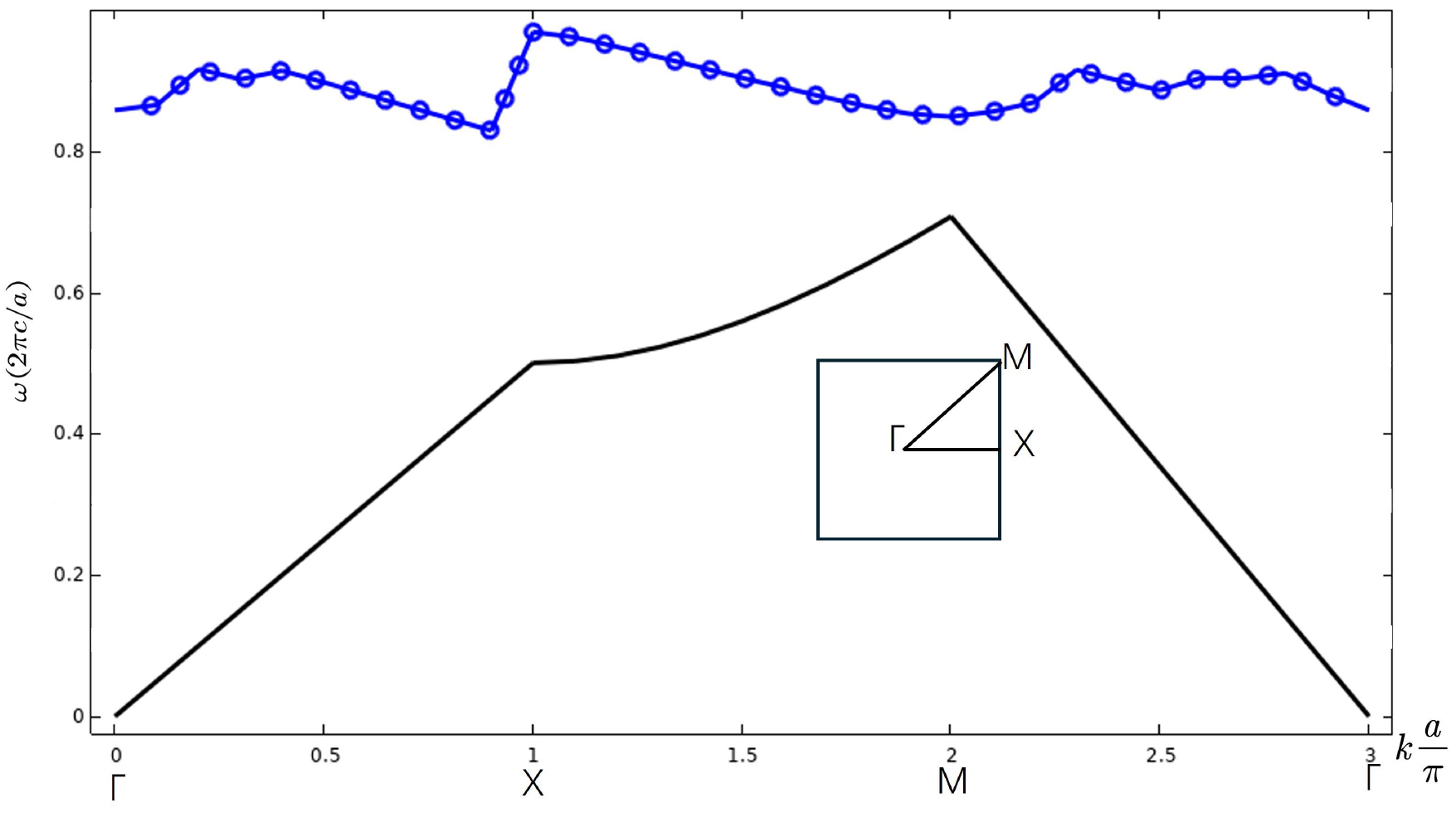}
		\label{fig:sub2}
	\end{subfigure}

		\begin{subfigure}[t]{0.23\textwidth}
		\captionsetup{justification=raggedright, singlelinecheck=false, labelformat=empty, skip=0pt, position=top}
		\caption*{(g)}
\includegraphics[height=3.2cm]{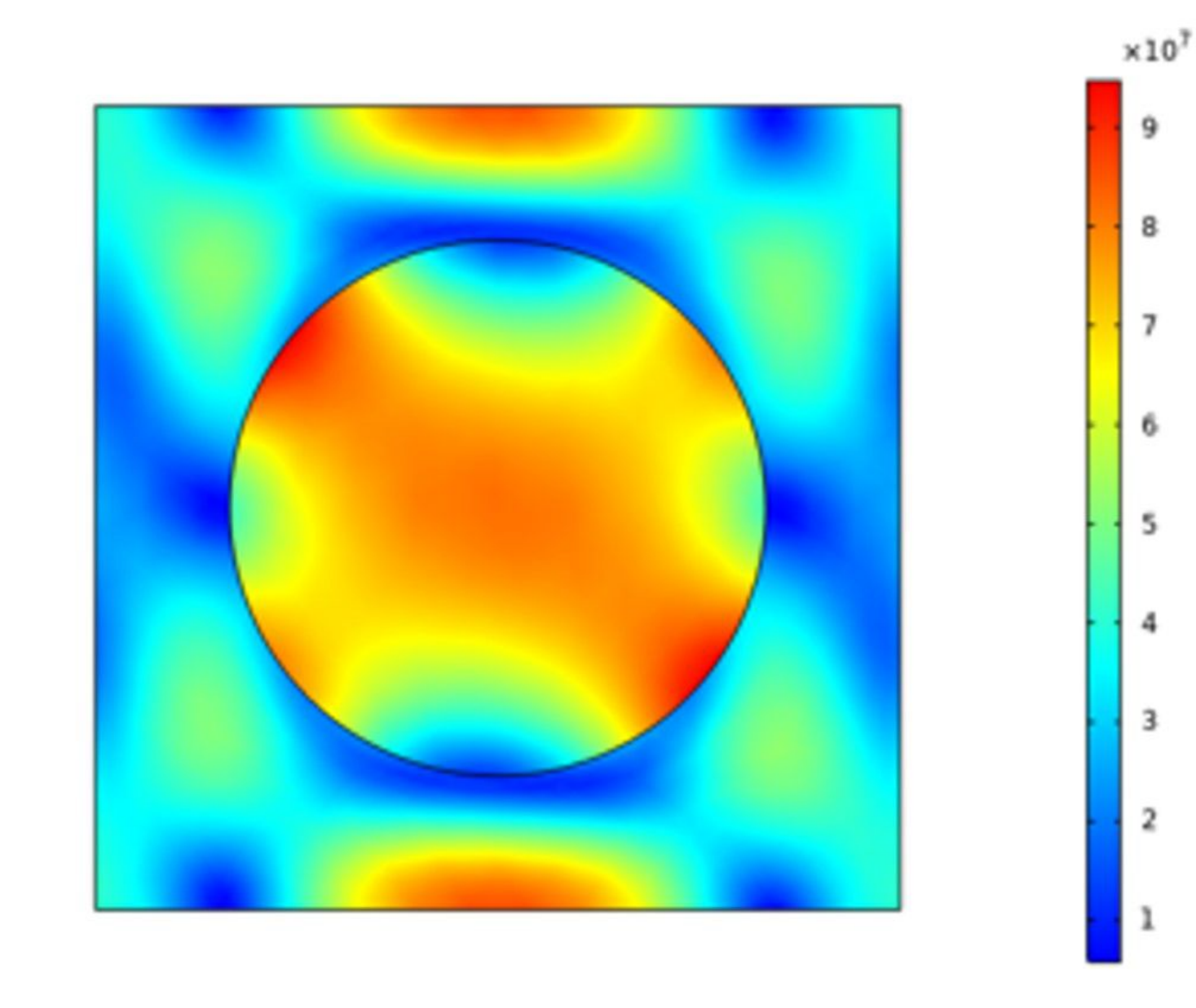}
		\label{fig:sub1}
	\end{subfigure}
	\begin{subfigure}[t]{0.32\textwidth}
		\captionsetup{justification=raggedright, singlelinecheck=false, labelformat=empty, skip=0pt, position=top}
		\caption*{(h)}
\includegraphics[height=3.2cm]{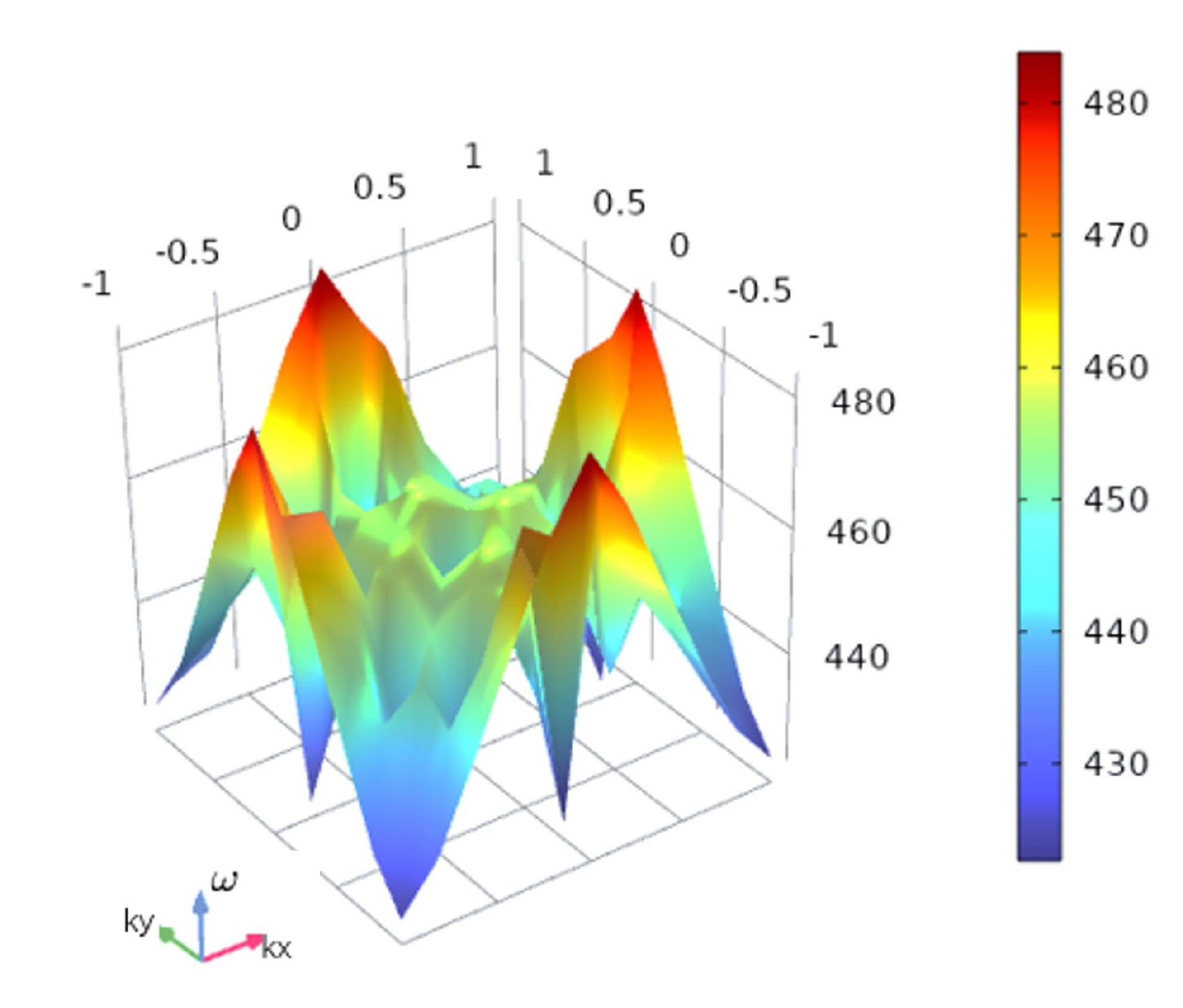}
		\label{fig:sub2}
	\end{subfigure}
		\begin{subfigure}[t]{0.23\textwidth}
		\captionsetup{justification=raggedright, singlelinecheck=false, labelformat=empty, skip=0pt, position=top}
		\caption*{(i)}
\includegraphics[height=3.2cm]{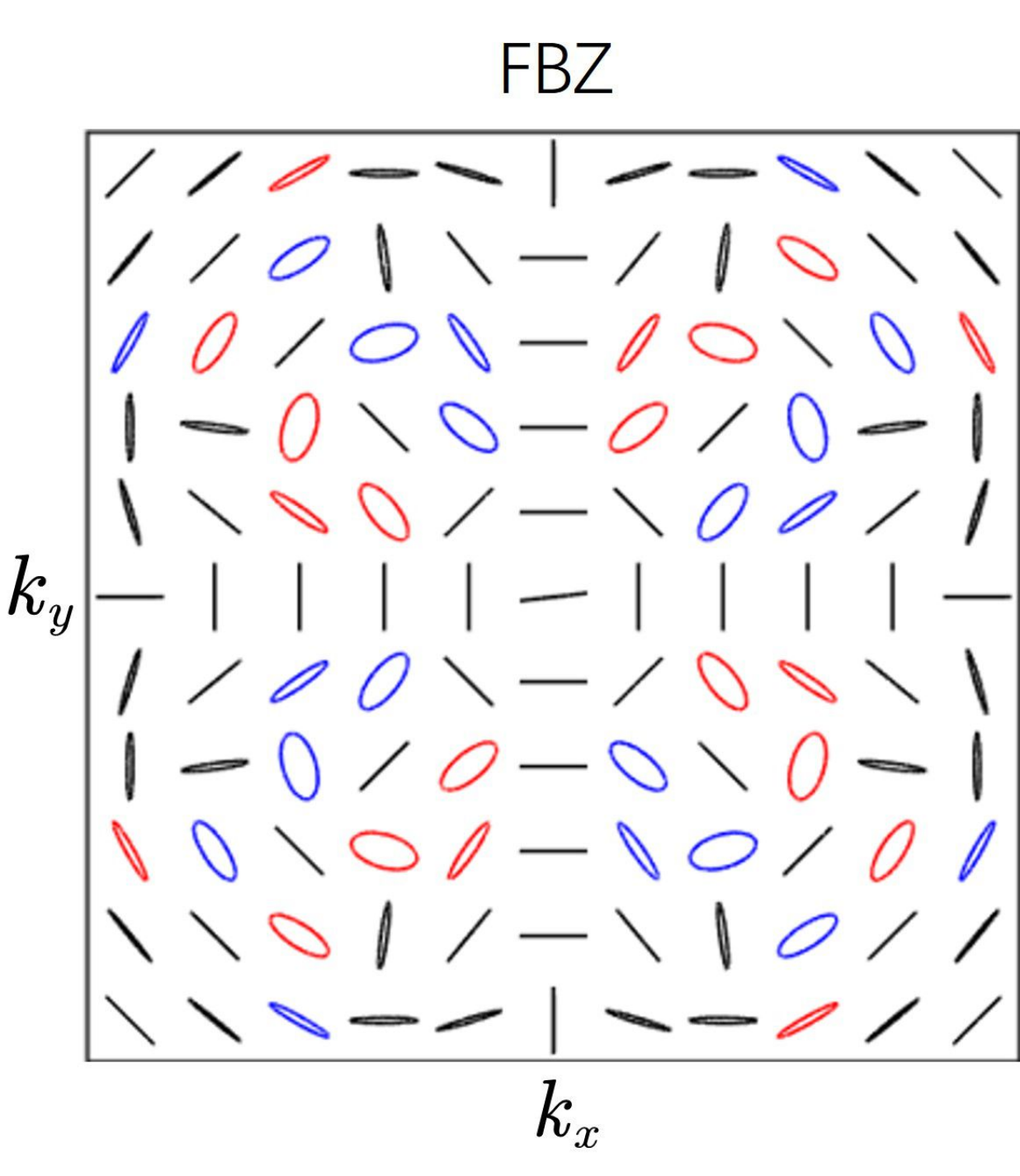}
		\label{fig:sub1}
	\end{subfigure}
	\hfill
	    
	        \captionsetup{justification=justified, singlelinecheck=false}

\caption{\label{fig:epsart}\justifying  Results of a square lattice slab.
(a) The energy band of the TE mode with a frequency of $342.42THz$ at the $\Gamma$ point. The black line represents the light cone, and the band is within the light cone.
(b) the magnetic field distribution at the $\Gamma$ point of (a).
(c) The complete band structure of (a) in the BZ.
(d) Distribution of the far-field polarization in the whole BZ for the band of (a).
(e) The quality factor of the BIC near the $\Gamma$ point.
(f) The energy band of the TE mode with a frequency of $428.61THz$ at the $\Gamma$ point. 
(g) the magnetic field distribution at the $\Gamma$ point for (f).
(h) The complete band structure of (f).
(i) Distribution of the far-field polarization for the band of (f).
}
\end{figure}

It should be clarified that the similarity transformation matrix in equation S10 comes from the following equation, which connects the eigenstates of far-field polarization and wavefunction of photonic crystals: 
\begin{equation}
\langle\varphi(k)|\hat{R}(\Lambda)| \varphi(k)\rangle=\left\langle C(k)\left|\Gamma^{*}(k) \hat{R}(\Lambda) \Gamma(k)\right| C(k)\right\rangle,
\end{equation}
where $\varphi(k)\rangle$ is the Bloch function: $|\varphi(k)\rangle=e^{i k r}|\phi(k)\rangle$. Due to the discrete translational invariance in the momentum space, we choose the $\hat{R}(\Lambda)$ to be the translation operation along the direction of BZ boundary:
\begin{equation}
\left\langle\varphi(k)\left|e^{-i K \hat{r} }\right| \varphi(k)\right\rangle=\left\langle C(k)\left|\Gamma^{*}(k) e^{-i K \hat{r}} \Gamma(k)\right| C(k)\right\rangle,
\end{equation}
where $K$ is the reciprocal lattice constant. we get the result below by expanding such equation:
\begin{equation}
\left\langle\varphi(k)\left|\sum_{n} \frac{(-i K)^{n}}{n!}\left(\nabla_{k}\right)^{n}\right| \varphi(k)\right\rangle=\left\langle C(k)\left|\Gamma^{*}(k) \sum_{n} \frac{(-i K)^{n}}{n!}\left(\nabla_{k}\right)^{n} \Gamma(k)\right| C(k)\right\rangle
\end{equation}

Since $K$ can be multiplied by an arbitrary integer, and the relation remains, each term on the left and right sides of it should be equal. Therefore,
\begin{equation}
\left\langle\phi(k)\left|\nabla_{k}\right| \phi(k)\right\rangle=\left\langle C(k)\left|\Gamma^{*}(k) \nabla_{k} \Gamma(k)\right| C(k)\right\rangle,
\end{equation}
which could be applied to equation S14.

In the following, we acquire the Chern number and winding number of far-field polarization states with specific crystal slabs, where the method in Ref.\nocite{r3} is applied.
Considering a square lattice photonic crystal slab with circular air holes with a radius of 200 nm on each lattice,  the refractive index of a medium is 2.5, the thickness is 200nm, and the lattice constant is 600nm. The eigenmode of TE mode with a frequency of $324.43THz$ at the $\Gamma$ point is analyzed, and its distribution of magnetic field in the z-component, together with the energy band, is shown in FIGs. S1a, S1b. The Chern number and winding are calculated to be -1, and a BIC with one topological charge exists at the $\Gamma$ point. For the band with a frequency of $428.61THz$ at the $\Gamma$ point, its chern number and winding number are -1 and 3, respectively. These results are all shown in FIG. S1.

\subsection{Calculation of Quality Factor near Gamma Point by Temporal Coupled-mode Theory}
During the topological phase transitions, it can be observed that when the gap closes, the band inversion accompanied by singular exchange occurs at the $\Gamma$ point, thereby altering the Chern number and the integral of the far-field polarization state. This section provides the topological charge of BIC at the $\Gamma$ point with the temporal coupled-mode theory \nocite{r6}. As shown in FIG. S2(a), a photonic crystal slab could be regarded as a resonant system with radiation leakage that radiates optic waves to the far field. According to \nocite{r7}, With the tight-binding model and the kp perturbation theory, a standard BHZ model or two QWZ models could be acquired, where the corresponding Hamiltonian is given by:
\begin{equation}
\hat{H}=\left(\begin{array}{cc}
\left(\begin{array}{cc}
w_{p}+\beta k^{2} & \alpha\left(k_{z}-i k_{v}\right) \\
\alpha\left(k_{x}+i k_{v}\right) & w_{d}-\beta k^{2}
\end{array}\right) & 0 \\
0 & \left(\begin{array}{cc}
w_{\mathrm{p}}+\beta k^{2} & \alpha\left(-k_{z}-i k_{v}\right) \\
\alpha\left(-k_{x}+i k_{v}\right) & w_{d}-\beta k^{2}
\end{array}\right)
\end{array}\right).
\end{equation}

Due to the presence of the pseudo-time-reversal symmetry in the system, half of the Hamiltonian could be treated as the resonant part concerning the temporal coupled-mode theory, and the following equations could be acquired:
\begin{equation}
-iw\begin{pmatrix}
    C_p^+\\C_d^+
\end{pmatrix}=-i\hat{H}^+\begin{pmatrix}
     C_p^+\\C_d^+
\end{pmatrix}-2\begin{pmatrix}
    \frac{1}{\tau_p} & 0\\
    0 & \frac{1}{\tau_d}
\end{pmatrix}\begin{pmatrix}
    C_p^+\\C_d^+
\end{pmatrix},
\end{equation}

\begin{equation}
\begin{pmatrix}
    E_p^+\\E_d^+
\end{pmatrix}_{\downarrow}=\begin{pmatrix}
    E_p^+\\E_d^+
\end{pmatrix}_{\uparrow}=\begin{pmatrix}
    \sqrt{\frac{2}{\tau_p}} & 0\\
    0 & \sqrt{\frac{2}{\tau_d}}
\end{pmatrix}\begin{pmatrix}
    C_p^+\\C_d^+
\end{pmatrix},
\end{equation}
where $\hat{H}^{+}=\left(\begin{array}{cc}
w_{p}+\beta k^{2} & \alpha\left(k_{x}-i k_{y}\right) \\
\alpha\left(k_{x}+i k_{g}\right) & w_{d}-\beta k^{2}
\end{array}\right)$, and $C_{p}^{+} ,\quad C_{d}^{+}$ is the magnitude of the component on $|p\rangle_{+},|d\rangle_{+}$, the factor 2 in the equation S20 is originated from two radiation directions of the leakage term, therefore,
\begin{equation}
\left|\begin{array}{cc}
w_{p}-i \frac{2}{\tau_{p}}-w & \alpha\left(k_{x}-i k_{y}\right) \\
\alpha\left(k_{x}+i k_{y}\right) & w_{d}-i \frac{2}{\tau_{d}}-w
\end{array}\right|=0.
\end{equation}

After expanding the equation S22, the following equation is obtained:
\begin{equation}
w^{2}-\left(w_{p}-i \frac{2}{\tau_{p}}+w_{d}-i \frac{2}{\tau_{d}}\right)+\left(w_{p}-i \frac{2}{\tau_{s}}\right)\left(w_{d}-i \frac{2}{\tau_{d}}\right)-\alpha^{2} k^{2}=0.
\end{equation}

Two modes are eventually obtained:
\begin{equation}
w_{ \pm}=\left(w_{p}-i \frac{2}{\tau_{p}}+w_{d}-i \frac{2}{\tau_{d}}\right) / 2 \pm \sqrt{\left(w_{p}-i \frac{2}{\tau_{p}}+w_{d}-i \frac{2}{\tau_{d}}\right)^{2}-4\left(w_{p}-i \frac{2}{\tau_{p}}\right)\left(w_{d}-i \frac{2}{\tau_{d}}\right)-\alpha^{2} k^{2}} / 2.
\end{equation}

Considering $k$ is small, and keeping the lowest-order term:

\begin{equation}
\begin{aligned}
w_+=w_p-i\frac{2}{\tau_p}+\frac{\alpha^2k^2}{w_p-i\frac{2}{\tau_p}-w_d+i\frac{2}{\tau_d}}\\
w_-=w_d-i\frac{2}{\tau_d}+\frac{\alpha^2k^2}{w_p-i\frac{2}{\tau_p}-w_d+i\frac{2}{\tau_d}}\\
\end{aligned}.
\end{equation}

According to Fig. 2b in the article, it can be observed that for the $|d\rangle_{+}$ mode, the z component of the magnetic field undergoes a sign change under inversion transformation, leading to a mismatched symmetry between the electric field inside the photonic crystal slab and the far field, leading to a BIC state. On the other hand, the $|p\rangle_{+}$ mode has the opposite parity compared to $|d\rangle_{+}$, causing strong coupling between the photonic crystal slab and far-field radiation. The decay time for the$|d\rangle_{+}$mode should tend to infinity, while for the$|p\rangle_{+}$mode, it becomes a finite value. The $w_-$ of Eq. S42 could be simplified as:
\begin{equation}
w_-=w_d-\frac{\alpha^2k^2}{a^2+b^2}(a+ib),
\end{equation}
where $a=w_{p}-w_{d}, \quad b=\frac{2}{\tau_{p}}$. 

If the term related to the square of $k$ in Eq. S22 is considered, $w_-$ would be $w_{d}-\left(\beta+\frac{\alpha^{2}}{a^{2}+b^{2}}(a+i b)\right) k^{2}$, and its quality factor is shown in FIG. S2(b), which is proportional to $\frac{1}{k^{2}}$. According to scale theory, a BIC charge with an absolute value of 1 exists near $\Gamma$ points.

	\begin{figure}[htbp]
	\centering
	\begin{subfigure}[t]{0.33\textwidth}
 \centering
		\captionsetup{justification=raggedright, singlelinecheck=false, labelformat=empty, skip=0pt, position=top}
		\caption*{(a)}
\includegraphics[height=3.2cm]{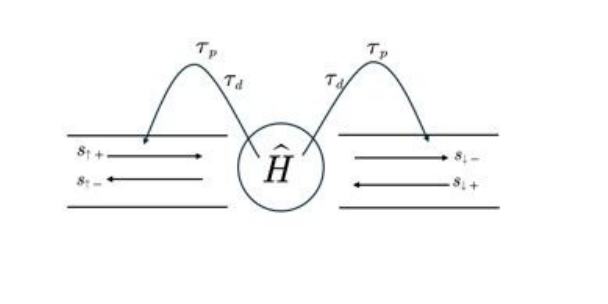}
		\label{fig:sub1}
	\end{subfigure}
	\begin{subfigure}[t]{0.34\textwidth}
		\captionsetup{justification=raggedright, singlelinecheck=false, labelformat=empty, skip=0pt, position=top}
		\caption*{(b)}
\includegraphics[height=3.2cm]{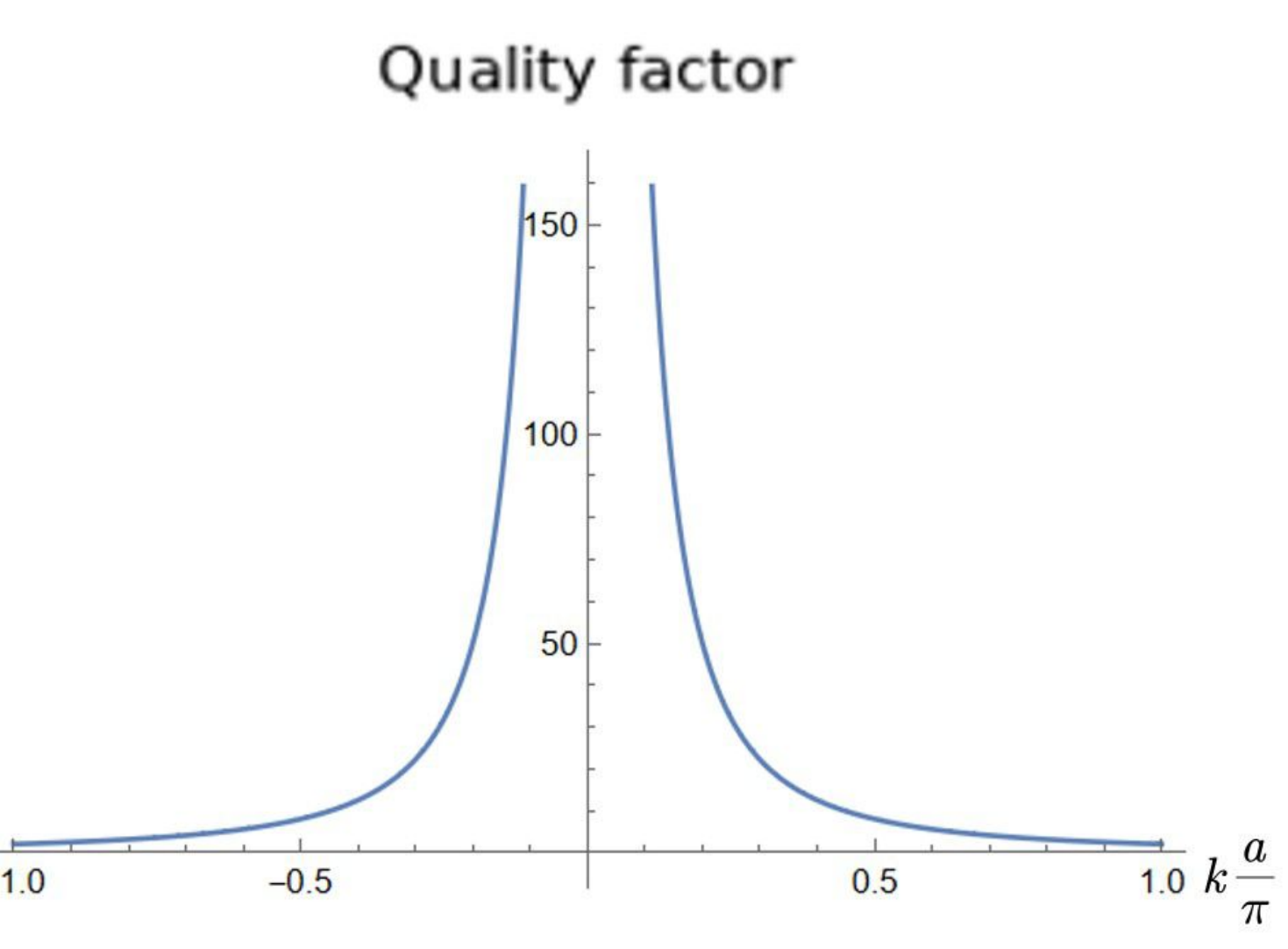}
		\label{fig:sub2}
	
	\end{subfigure}
	    
	        \captionsetup{justification=justified, singlelinecheck=false}

\caption{\label{fig:epsart}\justifying  Illustration of temporal coupled-mode theory.
(a) The resonant part of the photonic crystal slab is denoted with H, and there are two symmetric radiation channels, with two modes having a lifetime of in each channel.
(b) The quality factor distribution of the d-mode near the $\Gamma$ point, which is consistent with the results that a BIC with -1 charge exists at the $\Gamma$ point, as shown in Fig. 3c.
}
\end{figure}

\subsection{Robustness of Multi-layered Structures}
Previous sections suggested that the integral of the polarization angle around the boundary of the BZ is a topological invariant. This means that the integral will remain invariant as long as the energy gap remains open and the important symmetries of the Hamiltonian are unchanged under continuous parameter variations. We utilize a three-layer photonic crystal structure with equidistant spacing to provide further evidence for this conclusion. Energy splitting could be induced by varying the coupling strength between layers, and the integral of far-field polarization around the BZ remains unchanged.
The multi-layer structure is depicted in FIG. S3(a), consisting of three identical hexagonal photonic crystal layers. Its effective Hamiltonian is

\begin{equation}
\hat{H} =\sum_{d=1}^{D}|d\rangle\langle d|\otimes\hat{H}_{ld}+(C\sum_{d=1}^2|d+1\rangle\langle d|\otimes\hat{I}+c.c),
\end{equation}
where $\hat{H}_{ld}$ is the Hamiltonian of the $d$-th layer. We only consider the coupling between the neighbor layer here: the coupling will increase as the distance decreases. An appropriate gauge for |2> could also be determined so that $C$ becomes a real number.  The eigenstates of the three-layered structure could be used to get the eigenmode of the total Hamiltonian:

\begin{equation}
\phi=\sum_ic_i\varphi_i,
\end{equation}
where  $\varphi_i$ satisfies the relation $\hat{H}_{l d} \varphi_{i}=\varepsilon_{i}(k) \varphi_{i}$. The eigenvalue and eigenvectors of $\hat{H}$ could be acquired by following equations:

\begin{equation}
\begin{pmatrix}
    \hat{H}_{l1} & C\hat{I} & 0\\
    C^*\hat{I} & \hat{H}_{l2} & C\hat{I}\\
    0 & C^*\hat{I} & \hat{H}_{l3}
\end{pmatrix}\begin{pmatrix}
    c_1 \\ c_2 \\ c_3
\end{pmatrix}=E\begin{pmatrix}
    c_1 \\ c_2 \\ c_3
\end{pmatrix}
\end{equation}

\begin{equation}
E_{1}=\varepsilon(k) \quad E_{2}=\varepsilon(k)+\sqrt{2 C(k)} \quad E_{3}=E_{2}=\varepsilon(k)-\sqrt{2 C(k)}
\end{equation}

\begin{equation}
\phi_1=\begin{pmatrix}
    1\\0\\-1
\end{pmatrix}\hspace{0.2cm}\phi_2=\begin{pmatrix}
    1\\ \sqrt{\frac{2|C(k)|}{C^2(k)}} \\-1
\end{pmatrix}\hspace{0.2cm}\phi_1=\begin{pmatrix}
    1\\ -\sqrt{\frac{2|C(k)|}{C^2(k)}} \\1
\end{pmatrix}.
\end{equation}

	\begin{figure}[htbp]
	\centering
	\begin{subfigure}[t]{0.35\textwidth}
		\captionsetup{justification=raggedright, singlelinecheck=false, labelformat=empty, skip=0pt, position=top}
		\caption*{(a)}
\includegraphics[width=.95\linewidth]{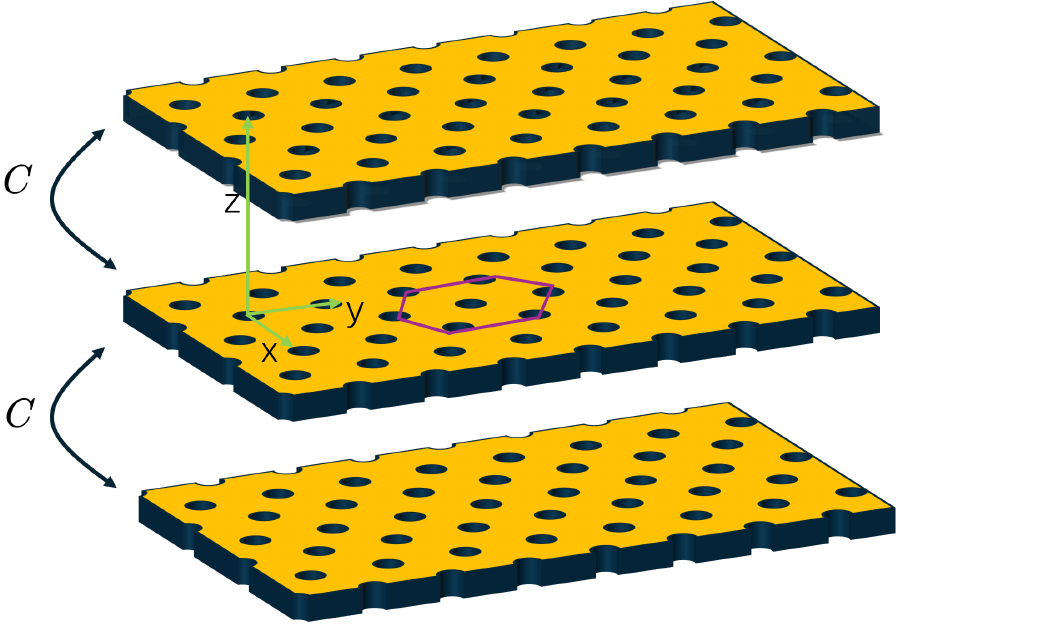}
		\label{fig:sub1}
	\end{subfigure}
	\begin{subfigure}[t]{0.85\textwidth}
		\captionsetup{justification=raggedright, singlelinecheck=false, labelformat=empty, skip=0pt, position=top}
		\caption*{(b)}
\includegraphics[width=.99\linewidth]{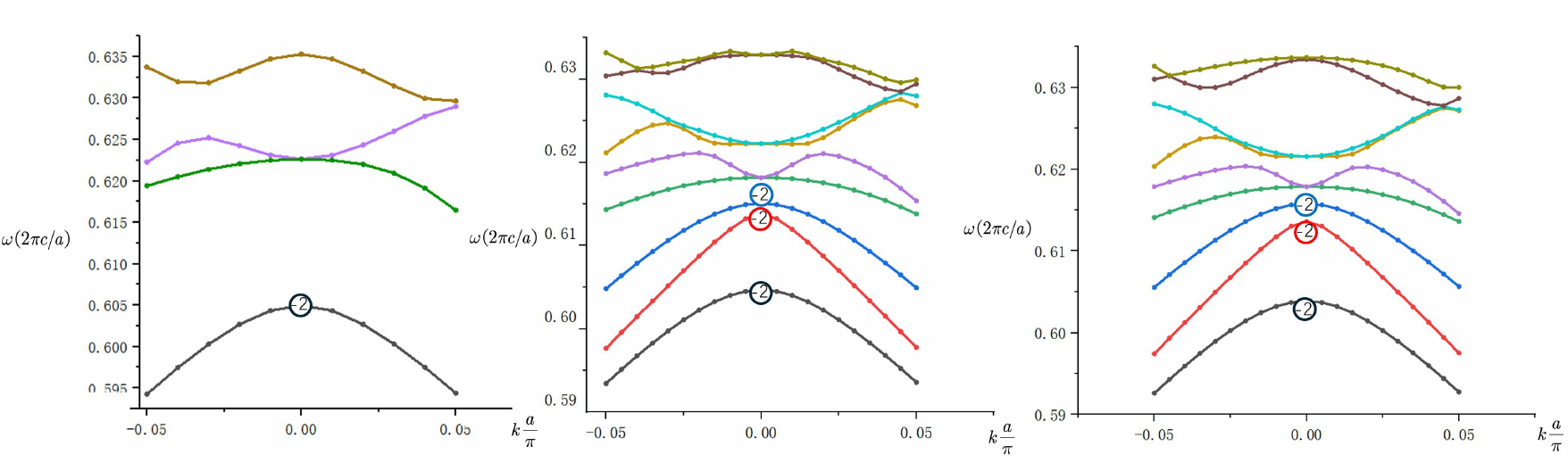}
		\label{fig:sub2}
	
	\end{subfigure}
	    
	        \captionsetup{justification=justified, singlelinecheck=false}

\caption{\label{fig:epsart}\justifying Multilayered structure and its band evolution.
(a) Multilayered structure. (b) Left: Energy bands when the distance between slabs is large; Middle: Energy bands for layers separated by 108nm; Right: Energy bands for layers separated by 98nm. It can be observed that during the evolution of the energy bands, the topological charge of BIC at the $\Gamma$ points remains unchanged.
}
\end{figure}

The evolution of energy bands is shown in FIG. S3(b). Due to the coupling between slabs, each energy band splits into multiple bands. The winding number of the red band corresponding to $\phi_2$ is calculated: 

\begin{equation}
\begin{array}{c}
q^{\prime}=\oint d k\left\langle c_{1} \varphi_{1}+c_{2} \varphi_{2}+c_{3} \varphi_{3}\left|i \nabla_{k}\right| c_{1} \varphi_{1}+c_{2} \varphi_{2}+c_{3} \varphi_{3}\right\rangle / 2 \pi \\
=\oint d k\left\langle c_{1} \varphi_{1}+c_{2} \varphi_{2}+c_{3} \varphi_{3} \mid\left(i \nabla_{k} c_{1}\right) \varphi_{1}+\left(i \nabla_{k} c_{2}\right) \varphi_{2}+\left(i \nabla_{k} c_{2}\right) \varphi_{2}\right\rangle / 2 \pi+ \\
\oint d k\left\langle c_{1} \varphi_{1}+c_{2} \varphi_{2}+c_{3} \varphi_{3} \mid\left(i \nabla_{k} \varphi_{1}\right) c_{1}+\left(i \nabla_{k} \varphi_{2}\right) c_{2}+\left(i \nabla_{k} \varphi_{3}\right) c_{3}\right\rangle / 2 \pi
\end{array}.
\end{equation}

Considering that $\nabla_{k}\left\langle\varphi_{i} \mid \varphi_{j}\right\rangle=0 \text { for } i \neq j$,

\begin{equation}
q^{\prime}=q_{2}+\oint d k \nabla_{k} \frac{1}{2} c_{2}{ }^{2}=q_{2}.
\end{equation}

The last term is a differential part; its integral over a closed loop is zero. Therefore, under continuous parameter changes, the winding number will remain unchanged. It should be noted that when a gap closes, the eigenvectors in equation S28 will not form closed subspaces because the close of the gap implies the presence of other eigenstates. Additionally, the form of the equation S29 will change when the symmetry of the Hamiltonian changes, potentially altering the winding number. 
\subsection{Far-field Effects of Edge States}

In this section, the edge states of the system are acquired by solving the Dirac equation, where the symmetry of their modes determines their quality factor. We only focus on the part $\hat{H}^{+}$ in s36 for systems with pseudo-time-reversal symmetry. Since there are periodic boundary conditions along the $x$ direction, $k_x$ is a good quantum quantity, and $k_y$ becomes$-i \frac{\partial}{\partial y}$due to the presence of the domain wall. For convenience, the center of the gap could be treated as the zero-energy reference, and the following Hamiltonian is obtained:

\begin{equation}
\hat{H}^+=\begin{pmatrix}
    u(y) & \alpha(k_x-ik_y)\\
    \alpha(k_x+ik_y) & -u(y)
\end{pmatrix}\hspace{0.5cm}k_y\xrightarrow{}-i\frac{\partial}{\partial y},
\end{equation}

Considering the dispersion relation near the $\Gamma$ point for the edge states is linear: $E(k)=\nu k_{x}$, the Schrödinger equation could be rewritten with the following form:
\begin{equation}
(u\hat{\sigma}_z+\alpha k_x\hat{\sigma }_x -i\alpha\frac{\partial}{\partial y}\hat{\sigma}_y)\varphi=\nu k_x\varphi.
\end{equation}

Multiplying $\hat{\sigma_y}$  to both sides of the equation and applying the relation that $\hat{\sigma}_{i} \hat{\sigma}_{j}=\delta_{i j}+i \epsilon_{i j k} \hat{\sigma}_{k}$, we know

\begin{equation}
(iu\hat{\sigma}_x-i\alpha k_x\hat{\sigma}_z-i\alpha\frac{\partial}{\partial y})\varphi=\nu k_x\hat{\sigma}_y\varphi,
\end{equation}
it can be seen that  $\varphi$ should be an eigenstate of $u\hat{\sigma}_x-\alpha k_x\hat{\sigma}_z+i\nu k_x\hat{\sigma}_y$:

\begin{equation}
(u\hat{\sigma}_x-\alpha k_x\hat{\sigma}_z+i\nu k_x\hat{\sigma}_y)\varphi_{\eta}=E_\eta\varphi_\eta,
\end{equation}
regarding the mass term $u$ in it is positively a constant in the region $y>0$, while it becomes $-u$ when $y<0$,  the magnitude of the magnetic field for the region $y<0$ is much larger. Therefore, we first calculate the eigenvalues for the region with negative mass: $E_{ \pm}= \pm \sqrt{\left(\alpha^{2}-\nu^{2}\right) k_{x}^{2}+u^{2}}$, and the respective eigenvector is

\begin{equation}
\varphi_\pm=\begin{pmatrix}
    1\\
    \frac{\nu k_x\pm\sqrt{(\alpha^2-\nu^2)k_x^2+u^2}}{u+\nu k_x}
\end{pmatrix}.
\end{equation}

Taking it into the equation S36, we obtain that $\alpha \frac{\partial}{\partial y} \varphi_{\eta}=E_{\eta} \varphi_{\eta}$, and the wave function of the state with negative energy is

\begin{equation}
\varphi=\begin{pmatrix}
    1\\
    \frac{\nu k_x-\sqrt{(\alpha^2-\nu^2)k_x^2+u^2}}{u+\nu k_x}
\end{pmatrix}e^{-\sqrt{(\alpha^2-\nu^2)k_x^2+u^2}|y|},
\end{equation}
where different parity may emerge when changing $k_x$: when $k_x$ equals to $-\frac{u}{\nu}$, edge states with high-quality factors exhibit the property $|d\rangle_{+} \propto d_{x^{2}-y^{2}}+i d_{x y}$ ; when $\nu k_{x} \pm \sqrt{\left(\alpha^{2}-\nu^{2}\right) k_{x}^{2}+u^{2}}=0$ , edge states with low-quality factors exhibit the property $|p\rangle_{+} \propto p_{x}+i p_{y}$ . FIG. S4(a) is the magnetic field distribution for the edge state in the topologically trivial region, where the theory is consistent with simulation. FIG. S4(b) shows the component of the d-mode in the edge state concerning $k_x$. It can be observed that the edge state closer to the lower energy band has a larger component of $cd+$ mode. This is consistent with Fig. 4d, where the quality factor also increases. 

	\begin{figure}[htbp]
	\centering
	\begin{subfigure}[t]{0.49\textwidth}
		\captionsetup{justification=raggedright, singlelinecheck=false, labelformat=empty, skip=0pt, position=top}
		\caption*{(a)}
\includegraphics[width=.95\linewidth]{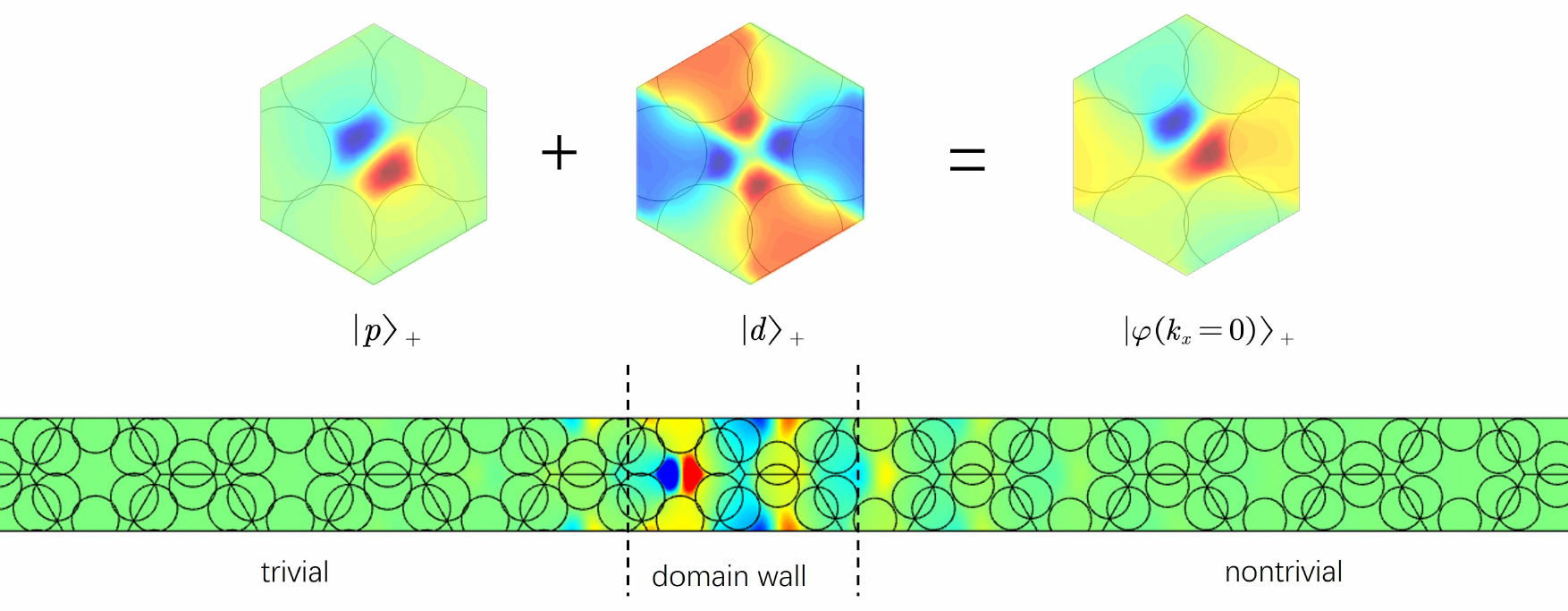}
		\label{fig:sub1}
	\end{subfigure}
	\begin{subfigure}[t]{0.49\textwidth}
		\captionsetup{justification=raggedright, singlelinecheck=false, labelformat=empty, skip=0pt, position=top}
		\caption*{(b)}
\includegraphics[width=.95\linewidth]{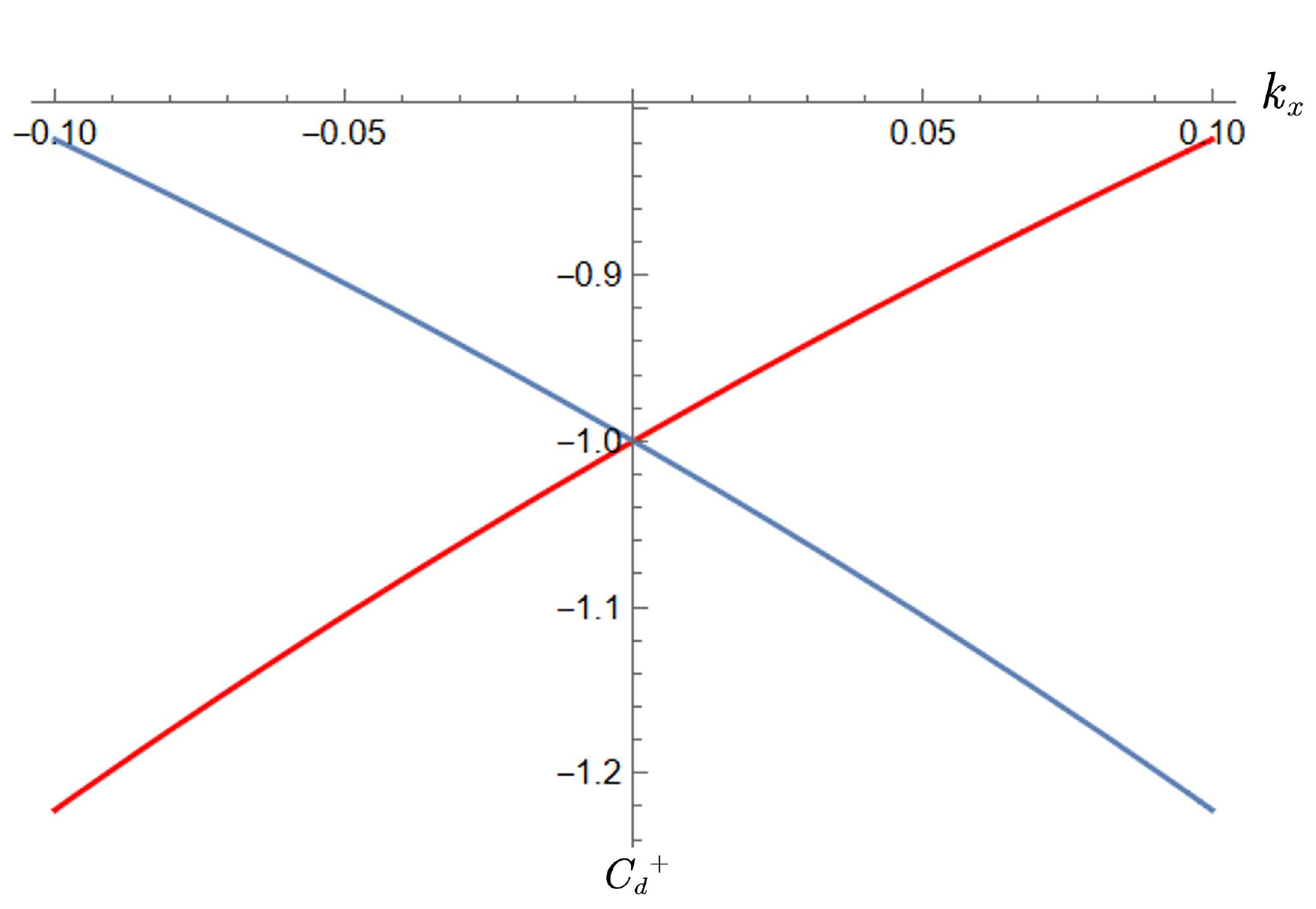}
		\label{fig:sub2}
	
	\end{subfigure}
	    
	        \captionsetup{justification=justified, singlelinecheck=false}

\caption{\label{fig:epsart}\justifying Diagram of edge states.
(a) Top: the superposition of the p-mode and d-mode at the $\Gamma$ point acquired from theory. Bottom: distribution of the edge state from simulation.
(b) Variation of the d-mode component of the edge state (red line) and its pseudo-time-reversal state (blue line) concerning $k_x$ in the topologically trivial region.
}
\end{figure}

\subsection{Constraints of Symmetry on Chern Numbers}

Previous studies on BICs have found that symmetry imposes strict restrictions on the values of their topological charges\nocite{r4}. After derivation, we discovered that for Chern topological insulators, like the two-dimensional photonic crystal with a Hamiltonian of $\hat{H}=\hat{d} \cdot \hat{\sigma}$, their Chern numbers are also constrained by symmetry.
Specifically, we adopt Kohmoto's method to calculate the Chern number: First, we need to define the reduced BZ for crystals with $C_{nv}$ symmetry, as shown in FIG. S5(a), where the reduced BZ for the hexagonal lattice is a triangle. When the system has a non-zero Chern number or lacks time-reversal symmetry, it is impossible to define a continuous gauge region in the BZ, indicating the existence of regions with different continuous gauges. We classify the distribution of regions with different continuous gauges into two cases.
Case 1: Existence of a Complete and Continuous Gauge Region in the reduced BZ as shown in FIG. S5(a), due to periodicity and symmetry, the reduced BZ of a hexagonal lattice can be folded and combined with its mirror-symmetric part, forming a closed geometric surface resembling two ice cream cones facing each other. Based on this, the integrated area for the Chern numbers will be divided into two parts, including S1 and S2:

\begin{equation}
C=\frac{1}{2\pi}\iint_{BZ}dk^2\Omega^z_{xy}=n\frac{1}{2\pi}\iint_{S_1}dk^2\Omega^z_{xy}+n\frac{1}{2\pi}\iint_{S_2}dk^2\Omega^z_{xy},
\end{equation}
where n represents the fold number of rotational symmetry the system possesses. With Stokes’ theorem, the Chern number could be obtained as below:
\begin{equation}
\begin{aligned}
C=\frac{1}{2 \pi} & \iint_{B Z} d k^{2} \Omega^{z}{ }_{x y}=n \frac{1}{4 \pi} \iint_{S_{1}} d k^{2} \Omega_{x y}^{z}+n \frac{1}{4 \pi} \iint_{S_{2}} d k^{2} \Omega_{x y}^{z} \\
& =n \frac{1}{4 \pi} \oint_{\partial s_{1}} \vec{A}_{1}(k) d k+n \frac{1}{4 \pi} \oint_{\partial s_{2}} \vec{A}_{2}(k) d k \\
& =n \frac{1}{4 \pi} \oint_{\partial s_{1}}\left(\vec{A}_{1}-\vec{A}_{2}\right) d k=n \frac{1}{4 \pi}\left(\gamma_{1}-\gamma_{2}\right),
\end{aligned}
\end{equation}
where $\vec{A}_{1}$ and $\vec{A}_{2}$ are the Berry connections in the S1 and S2 regions, respectively, and $\gamma_{1}$ ,$\gamma_{2}$ are the Berry phases for each region. Therefore, the Chern number can be expressed as $C = C_0 + nm$, where n is the fold number of the rotational symmetry of the system, and $m$ is $\frac{1}{4 \pi}\left(\gamma_{1}-\gamma_{2}\right)$, $C_0$ is determined by case 2 in the following.
Case 2: Existence of incomplete Gauge Regions in the reduced BZ.
 Now, let us calculate the value of $C_0$, where we need to transform the contour integral of a single-connected area into that of the area near high symmetrical points with the Kohmoto method. Since the calculation of the Berry connection involves the derivative of $\phi(k)$, which is a Bloch state, and it is a continuous function of $k$ and $r$, we can expand it as a Taylor series around the high-symmetry points in $k$-space:
\begin{equation}
\phi(k)=\sum_{nm}(a_{nm}+ib_{nm})k_+^mk_-^n,
\end{equation}
where $k_{+}=k_{x}+i k_{y}, k_{-}=k_{x}-i k_{y}$ . According to the equation S1 and the relationship between the reciprocal and real space and considering the wave function near high symmetrical points, we know that
\begin{equation}
\phi_{\Lambda k}(r)=\phi_k(\Lambda r), \phi(\Lambda^{-1}k)_n=\sum_mR_{nm}(\Lambda)\phi(k)_m.
\end{equation}

Taking the $B1$ representation of the $C_{6v}$ group as a example: $\phi\left(\sigma_{x} k\right)=-\phi(k), \quad \phi\left(\sigma_{y} k\right)=+\phi(k), \quad \phi\left(C^{z}_{6}{ }^{-1} k\right)=-\phi(k)$.By applying the these equation, we know that

\begin{equation}
\sum_{nm}(a_{nm}+ib_{nm})(-1)^{m+n}k^m_-k^n_+=-\sum_{nm}(a_{nm}+ib_{nm})k^m_+k^n_-,
\end{equation}
\begin{equation}
\sum_{nm}(a_{nm}+ib_{nm})k^m_-k^n_+=\sum_{nm}(a_{nm}+ib_{nm})k^m_+k^n_-,
\end{equation}
\begin{equation}
\sum_{nm}(a_{nm}+ib_{nm})e^{-i(m-n)\frac{\pi}{3}}k^m_+k^n_-=-\sum_{nm}(a_{nm}+ib_{nm})k^m_+k^n_-,
\end{equation}
from which we know that $a_{nm}=a_{mn}$, $b_{nm}=b_{mn}$, $m+n$ is an odd number. According to equation S46, $n-m=3N$, where $N$ is an integer. Taking the state $\phi\left(k_{x}, k_{y}=0\right)$ and the rotated state $\phi\left(k_{x}, k_{y}=\frac{\sqrt{3}}{2} k_{x}\right)$ into equation S46,

\begin{equation}
\sum_{nm}a_{mn}k^m_+k^n_-=\sum_{nm}(\frac{1}{2}a_{nm}+\frac{\sqrt{3}}{2}b_{nm})k^m_+k^n_-.
\end{equation}

So $b_{m n}=\frac{\sqrt{3}}{3} a_{m n}$ , and $\phi_{k}(r)=C_{1}(r)\left(k_{+}^{3}+k_{-}^{3}\right)+O\left(k^{6}\right)$ near high symmetrical points.
Now lets calculate the value of $\oint_{\partial \Delta} i\left\langle\phi_{k}(r)\left|\nabla_{k}\right| \phi_{k}(r)\right\rangle d k / 2 \pi$, namely $C_0$, where $\Delta$ is the continuous gauged regions near high symmetrical points, and $\partial \Delta$ is the corresponding boundary:
\begin{equation}
\begin{array}{c}
C_{0}=\oint_{\partial \Delta} i\left\langle\phi_{k}(r)\left|\nabla_{k}\right| \phi_{k}(r)\right\rangle d k / 2 \pi=\oint_{\partial \Delta}\left(k_{+}^{3}+k_{-}^{3}\right)^{*} \nabla_{k}\left(k_{+}^{3}+k_{-}^{3}\right) d k / 2 \pi \\
=\oint_{\partial \Delta} \frac{1}{2} \nabla_{k}\left(k_{+}^{3}+k_{-}^{3}\right)^{2} d k / 2 \pi=0.
\end{array}
\end{equation}
The high symmetrical points enclosed by the reduced BZ for square and hexagonal lattices are shown in FIG. S5(b).

	\begin{figure}[htbp]
	\centering
	\begin{subfigure}[t]{0.53\textwidth}
		\captionsetup{justification=raggedright, singlelinecheck=false, labelformat=empty, skip=0pt, position=top}
		\caption*{(a)}
\includegraphics[width=.95\linewidth]{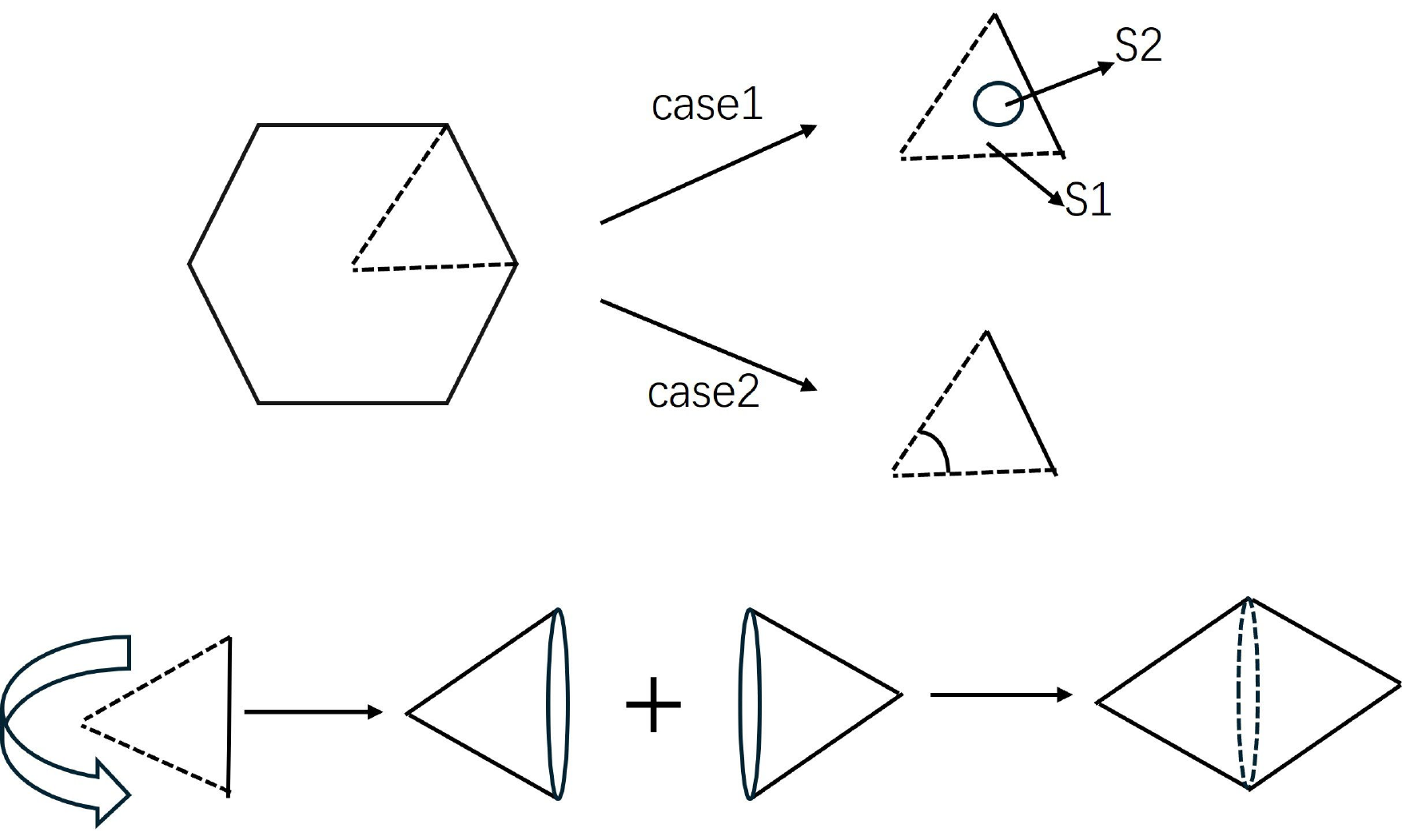}
		\label{fig:sub1}
	\end{subfigure}
 
	\begin{subfigure}[t]{0.55\textwidth}
		\captionsetup{justification=raggedright, singlelinecheck=false, labelformat=empty, skip=0pt, position=top}
		\caption*{(b)}
\includegraphics[width=.95\linewidth]{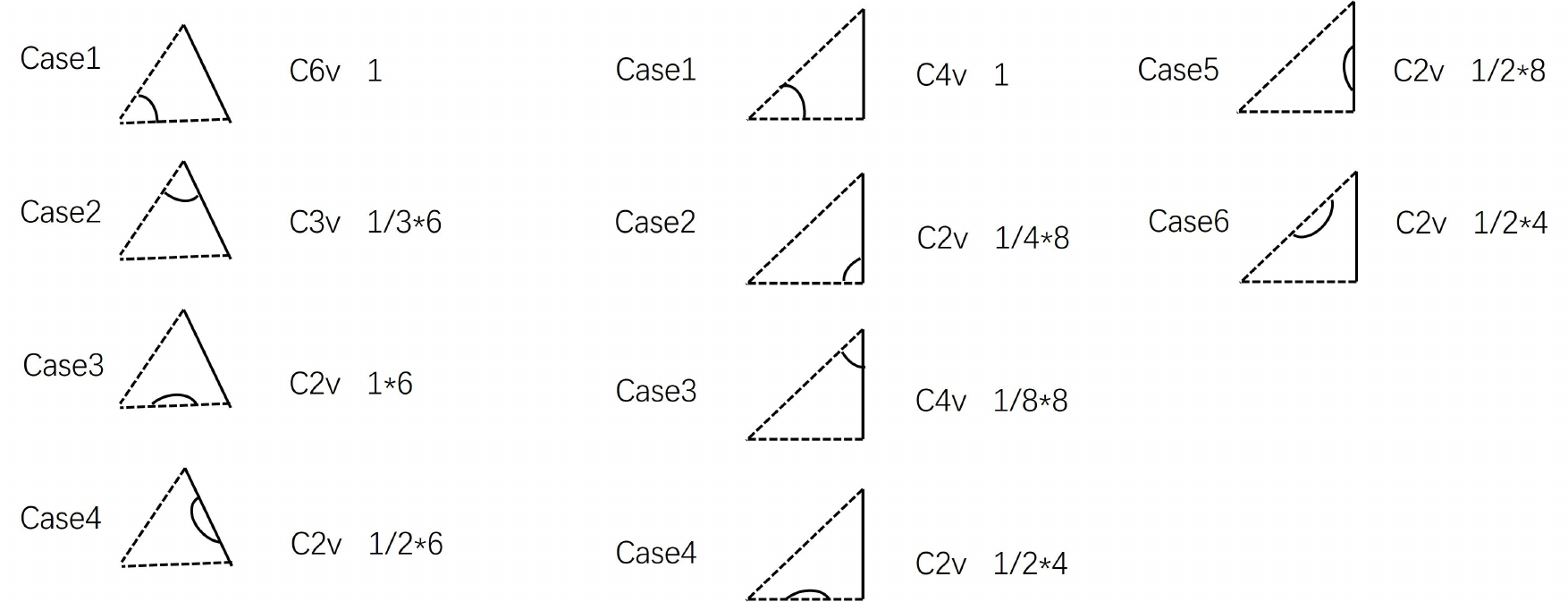}
		\label{fig:sub2}
	
	\end{subfigure}
	    
	        \captionsetup{justification=justified, singlelinecheck=false}

\caption{\label{fig:epsart}\justifying 
(a) The distribution of the continuous gauge regions in the reduced BZ of hexagonal lattices.
(b) The distribution of complete and incomplete regions in the reduced BZ for hexagonal and square lattices.
}
\end{figure}

\subsection{Correspondence under non-Abelian connections}
This section will discuss how to identify topological correspondence in the presence of non-Abelian connections. To begin with, one-dimensional crystal slabs are considered, where the Wilson loop will be a matrix: 

\begin{equation}
W=M^{(12)}M^{(23)}\dots M^{(N-1,N)}M^{N1},
\end{equation}
where $N_f$ is the degeneracy of energy bands, and $M^{i j}$ is a $N_{f} \times N_{f}$ matrix:$M_{n m}{ }^{i j}=\left\langle\phi_{n}\left(k_{i}\right) \mid \phi_{n}\left(k_{j}\right)\right\rangle$ Therefore, $W$ is a gauged invariant, and its eigenvalue is related to the position of the Wannier center or crystal polarization\nocite{r5}. To acquire the eigenvalue, we define a projection operator: $\hat{P}=\sum_{k} \sum_{n=1}^{N_{f}}\left|\varphi_{n}(k)\right\rangle\left\langle\varphi_{n}(k)\right|$, where $\varphi_n(k)\rangle$ is Bloch states:

\begin{equation}
|\varphi_n(k)\rangle=\frac{1}{\sqrt{N}}\sum_{m=1}^Ne^{imk}|m\rangle\otimes|\phi_n(k)\rangle.
\end{equation}
For convenience, we define the lattice constant $a$ as 1, and $\delta k=\frac{2 \pi}{N}$, the Wannier state for bands with degeneracy becomes:
\begin{equation}
|w_n(j)\rangle=\frac{1}{\sqrt{N}}\sum_{k=\delta_k}^{N\delta_k}e^{-ijk}\sum_{p=1}^{N_f}U_{np}(k)|\varphi_p(k)\rangle,
\end{equation}
,where $U$ is dependent on $k$. Therefore, the position of the $j$th Wannier state of the nth energy band can be obtained:
\begin{equation}
\langle x\rangle_{n,j}=\frac{N}{2\pi}arg\langle w_n(j)|\hat{X}_p|w_n(j)\rangle=\frac{N}{2\pi}arg\lambda_{n,j}=\langle x\rangle_n+j,
\end{equation}
where the $\hat{X}_p$ is satisfies $\left(\hat{X}_{p}\right)^{N}=\sum_{k} \otimes \sum_{m n}^{N_{f}} W_{m n}{ }^{(k)}\left|\varphi_{m}(k)\right\rangle\left\langle\varphi_{n}(k)\right|
$, and $\left(X_{p}\right)^{N}$ is a $(N*N_f)\times (N*N_f)$ dimensional matrix. The eigenvalue of $X_{p}$ is naturally $\lambda_{n, j}=e^{i \theta_{n} / N+i j \delta k+\ln \left(\left|\lambda_{n}\right|\right) / N}$. For the equation S51, one can define:
\begin{equation}
|\phi_n'(k)\rangle=\sum_{p=1}^{N_f}U_{np}|\phi_p(k)\rangle.
\end{equation}

The polarization state for the n-th bands could be defined as $P_{n}=\frac{i}{2 \pi} \int d k\left\langle\phi_{n}{ }^{\prime}(k)\left|\frac{\partial}{\partial k}\right| \phi_{n}^{\prime}(k)\right\rangle$, therefore, the pumping number of Wannier center for photonic crystal plates could be defined:
\begin{equation}
Pum_n=\frac{i}{2\pi}\oint dk\langle\phi_n'(k)|\nabla_k|\phi_n'(k)\rangle,
\end{equation}
and consider equation S9, we acquired that:
\begin{equation}
|\phi_n'(k)\rangle=\sum_{q,p}\Gamma_{nq}U_{qp}|C_p(k)\rangle.
\end{equation}

Since the multiplication of two unitary matrices is also unitary: $(A B)^{+} A B=B^{+} A^{+} A B=I$, we define a unitary transformation $R_{n m}=\sum_{q} \Gamma_{n q} U_{q m}$, and take equation S58 into the equation S57:
\begin{equation}
\begin{aligned}
\operatorname{Pum}_{n} & =\frac{i}{2 \pi} \oint d k\left\langle\phi_{n}{ }^{\prime}(k)\left|\nabla_{k}\right| \phi_{n}{ }^{\prime}(k)\right\rangle=\oint \sum_{i, j} R_{n i}^{+}\left\langle C_{i}(k)\left|i \nabla_{k}\right| C_{j}(k)\right\rangle R_{j n} d k \\
& =\oint d k \sum_{i j}\left[\left\langle C_{i}(k)\left|i \nabla_{k}\right| C_{j}(k)\right\rangle\right] R_{j n} R_{n i}^{+}+\oint d k \sum_{i j} R_{n i}^{+} i \nabla_{k} R_{j n}.
\end{aligned}
\end{equation}

Since the elements of unitary matrices in single-valued, the last term of it should be a constant. A mapping connection could be acquired by summing over n on both sides of the equation S59:
\begin{equation}
\begin{aligned}
\sum_{n} P{u m_n} & =\oint d k \sum_{i, j, n} R_{j n} R_{n i}^{\dagger}\left\langle C_{i}(k)\left|i \nabla_{k}\right| C_{j}(k)\right\rangle+C \\
& =\sum_{i} q_{i}+C
\end{aligned}
\end{equation}

Therefore, a mapping connection exists between the sum of the charge pumping number and the winding number for the far-field polarization.

\end{singlespacing}



\end{document}